\documentclass[useAMS,usenatbib]{mn2e}

\usepackage{graphicx} 

\usepackage{amsmath}

\title[Unifying X-ray winds in Seyfert galaxies]{Unification of X-ray winds in Seyfert galaxies: from ultra-fast outflows to warm absorbers}
\author[F. Tombesi et al.]{F. Tombesi$^{1,2}$\thanks{E-mail: ftombesi@astro.umd.edu}, M. Cappi$^{3}$, J.~N. Reeves$^{4}$, R.~S. Nemmen$^1$, V. Braito$^{5}$, M. Gaspari$^{6}$ \newauthor and C.~S. Reynolds$^2$\\
$^{1}$NASA Goddard Space Flight Center, Greenbelt, MD 20771, USA\\
$^{2}$Department of Astronomy, University of Maryland, College Park, MD 20742, USA\\
$^3$INAF-IASF Bologna, Via Gobetti 101, I-40129 Bologna, Italy\\
$^{4}$Astrophysics Group, School of Physical and Geographical Sciences, Keele University, Keele, Staffordshire ST5 5BG, UK\\
$^5$INAF-Osservatorio Astronomico di Brera, via. E. Bianchi 46, I-23807, Merate, Italy\\
$^{6}$Max Planck Institute for Astrophysics, Karl-Schwarzschild-Strasse 1, 85741 Garching, Germany
}
\begin{document}



\maketitle

\label{firstpage}

\begin{abstract}

The existence of ionized X-ray absorbing layers of gas along the line of sight to the nuclei of Seyfert galaxies is a well established observational fact. This material is systematically outflowing and shows a large range in parameters. However, its actual nature and dynamics are still not clear. In order to gain insights into these important issues we performed a literature search for papers reporting the parameters of the soft X-ray warm absorbers (WAs) in 35 type 1 Seyferts and compared their properties to those of the ultra-fast outflows (UFOs) detected in the same sample. The fraction of sources with WAs is $>$60\%, consistent with previous studies. The fraction of sources with UFOs is $>$34\%, $>$67\% of which also show WAs. The large dynamic range obtained when considering all the absorbers together, spanning several orders of magnitude in ionization, column, velocity, and distance allows us, for the first time, to investigate general relations among them. In particular, we find significant correlations indicating that the closer the absorber is to the central black hole, the higher the ionization, column, outflow velocity and consequently the mechanical power.  
In all the cases, the absorbers continuously populate the whole parameter space, with the WAs and the UFOs lying always at the two ends of the distribution. These evidences strongly suggest that these absorbers, often considered of different types, could actually represent parts of a single large-scale stratified outflow observed at different locations from the black hole. The UFOs are likely launched from the inner accretion disc and the WAs at larger distances, such as the outer disc and/or torus. 
We argue that the observed parameters and correlations are, to date, consistent with both radiation pressure through Compton scattering and MHD processes contributing to the outflow acceleration, the latter playing a major role. Most of the absorbers, especially the UFOs, show a sufficiently high mechanical power (at least $\sim$0.5\% of the bolometric luminosity) to provide a significant contribution to AGN feedback and thus to the evolution of the host galaxy. In this regard, we find possible evidences for the interaction of the AGN wind with the surrounding environment on large-scales.

\end{abstract}

\begin{keywords}
accretion, accretion discs -- black hole physics -- galaxies: active -- galaxies: Seyfert -- X-rays: galaxies
\end{keywords}

\section{Introduction}

The presence of ionized material along the line of sight to the nuclei of Seyfert galaxies has been known for a long time (e.g. Halpern 1984). The material observable in absorption in the soft X-rays has been referred to as a warm absorber (WA). The limited energy resolution of previous X-ray instruments allowed essentially only the detection of broad absorption edges. In fact, observations with ASCA (Advanced Satellite for Cosmology and Astrophysics) detected O~VII and O~VIII absorption edges and established that WAs are a common feature of AGNs, being present in $\sim$50\% of Seyfert galaxies (Reynolds 1997; George et al. 1998). Then, the advent of the unprecedented spectral resolution of the gratings on board \emph{Chandra} and \emph{XMM-Newton} allowed, for the first time, the detection of discrete soft X-ray resonance absorption and emission lines. The resulting picture of the WA is that of an outflow exhibiting multiple narrow absorption lines corresponding to different ionization states (Kaastra et al.~2000; Kaspi et al.~2000; Blustin et al.~2005; McKernan et al.~2007). The values of the ionization parameter are typically in the range log$\xi$$\sim$0--2~erg~s$^{-1}$~cm and the column densities are between $N_H$$\sim$$10^{20}$--$10^{22}$~cm$^{-2}$. The absorption lines are systematically blue-shifted, indicating outflow velocities of the WAs in the range $v_\mathrm{out}$$\sim$100--1000~km/s. There are still significant uncertainties on the exact location of this material, which ranges from $\sim$pc up to $\sim$kpc scales, and it has been suggested that it might originate outside of the inner disc, probably at locations comparable with the obscuring torus (e.g. Krolik \& Kriss 2001; Blustin et al.~2005). Depending on the actual filling and covering factors, the mass outflow rate from the WAs can be significant but, given their relatively low velocities, their kinetic power is rather low when compared to the bolometric luminosity (e.g. Blustin et al.~2005; McKernan et al~2007). However, a recent detailed study by Crenshaw \& Kraemer (2012) found that the integrated power of the WAs in some Seyferts can actually reach the level of $\sim$0.1--0.5\% of the bolometric luminosity, the minimum required by numerical simulations for AGN feedback to exert a significant impact on the host galaxy (e.g. Hopkins \& Elvis 2010; Gaspari et al.~2011a, b, 2012b).  

Besides WAs, highly blueshifted Fe K-shell absorption lines at E$\ga$7~keV have been detected in more recent years in the X-ray spectra of several AGNs (Chartas et al. 2002, 2003; Pounds et al.~2003; Dadina et al.~2005; Markowitz et al.~2006; Braito et al.~2007; Cappi et al.~2009; Reeves et al.~2009; Chartas et al.~2009; Giustini et al.~2011; Gofford et al.~2011; Lobban et al.~2011; Dauser et al.~2012). In particular, a uniform and systematic search for blueshifted Fe K absorption lines in a sample of 42 local ($z \le 0.1$) Seyferts observed with \emph{XMM-Newton} was performed by Tombesi et al.~(2010a, hereafter Paper I). This allowed the authors to assess their global significance and derive a detection fraction of $\ga$40\%. 
In order to mark an initial and somewhat arbitrary distinction with the classical WAs, in paper I we defined ultra-fast outflows (UFOs) as those highly ionized Fe K absorbers with a blueshifted velocity $\ge$10,000~km/s. Subsequently, Tombesi et al.~(2011a, hereafter Paper II) performed a photo-ionization modelling of these UFOs and derived the distribution of their main physical parameters. The outflow velocity is mildly-relativistic, in the range $\sim$0.03--0.3c ($\sim$10,000--100,000~km~s$^{-1}$), with mean value of $\sim$0.1c. The ionization is very high, in the range log$\xi$$\sim$3--6~erg~s$^{-1}$~cm, with a mean value of $\sim$4.2~erg~s$^{-1}$~cm. The column densities are also large, in the interval $N_H$$\sim$$10^{22}$--$10^{24}$~cm$^{-2}$. These findings are important because they suggest the presence of massive and highly ionized absorbers outflowing with mildly-relativistic velocities from the nuclei of these Seyfert galaxies.

In a following paper, Tombesi et al.~(2012a, hereafter Paper III) quantified that they are observable at locations of sub-parsec scales from the central black hole. Their mass outflow rate was constrained between $\sim$0.01--1~$M_{\odot}$~yr$^{-1}$ and their kinetic power was found in the range log$\dot{E}_\mathrm{K}$$\simeq$42--45~erg~s$^{-1}$. Thus, the UFOs are possibly directly identifiable, at least qualitatively, with accretion disc winds/outflows (King \& Pounds 2003; Proga \& Kallman 2004; Ohsuga et al.~2009; Sim et al.~2010; Fukumura et al.~2010) or the base of a possible weak jet (e.g. Ghisellini et al.~2004). In particular, the kinetic power of these UFOs is systematically higher than the minimum required by simulations of feedback induced by winds/outflows (e.g. Hopkins \& Elvis 2010; Gaspari et al.~2011a, b, 2012b). Therefore, in the long term, they could be able to significantly influence the bulge evolution, star formation, super-massive black hole growth and contribute to the establishment of the observed black hole-host galaxy relations, such as the $M_{BH}$--$\sigma$ (Ferrarese \& Merritt 2000; King 2010a; Ostriker et al.~2010; Gaspari et al.~2011a, b, 2012a, b).

It is important to note that Tombesi et al.~(2010b, 2011b, 2012b) detected the presence of UFOs also in a small sample of (3 out of 5) radio-loud AGNs observed with \emph{Suzaku}. Finally, we also note that similar results regarding the UFOs have been obtained independently by Gofford et al.~(2012) who performed a uniform and systematic broad-band spectral analysis of a large sample of AGNs observed with \emph{Suzaku}, confirming their overall incidence and characteristics. Moreover, evidences for similar outflows are emerging also in stellar-mass black holes (e.g., King et al.~2012)

Despite these significant observational developments, the origin and acceleration mechanism(s) of the ionized absorbers in AGNs are still debated. In particular, radiation and MHD wind models have been developed and scenarios in which they are considered as intrinsically distinct or as different manifestations of the same phenomenon have been suggested (e.g. K\"{o}nigl \& Kartje 1994; Krolik \& Kriss 2001; Elvis 2000; Blustin et al.~2005;  Krongold et al.~2007; Fukumura et al.~2010; Kazanas et al.~2012; Reynolds 2012). Given the relevance of these outflows for the physics and energetics of AGNs and their potential significant contribution to feedback, it is imperative to investigate them in detail.

In order to test these hypothesis, in this paper we will perform a detailed observational comparison of the WAs and UFOs, checking for correlations and discussing the possible unification of these absorbers in a single, photo-ionized and stratified outflow. We focus on the sample of Seyfert galaxies described in Paper I and we will use WA parameters collected from the literature and those of the UFOs derived in Paper II and Paper III, respectively.

\section[]{Warm absorbers selection}

We performed a literature search for papers reporting the analysis of the soft X-ray WAs in the 35 type 1 Seyferts of the sample discussed in Paper I. This was defined selecting all the Seyfert galaxies from the \emph{RXTE} All-Sky Slew Survey Catalog (Revnivtsev et al. 2004) and cross-correlating them with the \emph{XMM-Newton} Accepted Targets Catalog (as of October 2008). After applying the standard filtering processes, we obtained a total of 42 objects with 101 good {\emph XMM-Newton} observations and, more specifically, 35 classified as type 1 and 7 as type 2 with 87 and 14 observations, respectively.

In the literature search, we selected only the WA results derived from the high energy resolution gratings on board \emph{Chandra} and \emph{XMM-Newton} because they allow us to constrain the outflow velocity, which is crucial to derive the mass outflow rate and the kinetic power. We limited our search only to the type 1 sources because, in accordance with the unification model, the spectra of the 7 type 2s are affected by significant neutral absorption ($N_{\mathrm{H}}$$\ga$$10^{23}$~cm$^{-2}$) which hampers the detection of the WAs in the soft X-rays. For the UFOs, we consider the parameters reported in Paper II and III. 

In the following, we use the definition of the ionization parameter $\xi \equiv L_\mathrm{ion}/nr^2$ (Tarter et al.~1969) in which $L_\mathrm{ion}$ is the ionizing luminosity between 1~Ryd and 1000~Ryd (1~Ryd $=$ 13.6~eV), $n$ is the number density of the absorbing material and $r$ is the distance from the central source.
Often, different WA components with diverse ionization, velocities and column densities are detected for each source and there might be time variability, especially between observations spaced by several years. Moreover, there might be some intrinsic inhomogeneities in the density and ionization structure of the absorbing material. However, here we do not consider any subtle variations in the warm absorber properties because we focus on deriving the global properties of the outflows and we refer the reader to the papers reported in the notes of Table~1 for more information. Therefore, for each source we report the various ionized absorption components in Table~1. If more than one paper reported the analysis of the same source, we averaged the values of the components with equivalent parameters.
This allows us to minimize the scatter due to different analysis methods employed by different authors and the effects of time variability as well.  The WA parameters for all the sources, along with their central black hole masses and average (absorption corrected) ionizing luminosities $L_\mathrm{ion}$ derived from the \emph{XMM-Newton} observations reported in Paper I are reported in Table~1. In the subsequent correlation analysis we will include the points from these multiple zones of warm absorbers separately.   

For comparison, we use the parameters of the highly ionized Fe K-shell absorbers reported in Paper II. These were initially simply distinguished as UFOs or non-UFOs if their velocity was higher or lower than 10,000~km/s, respectively. However, as noted in Paper II and III, their parameter distributions are not bimodal, but they actually show a roughly continuous distribution in ionization, column density and velocity, with the UFOs lying at the more extreme side. This point is also confirmed by the analysis of Gofford et al.~(2012). Bearing this in mind, in the following we will continue to refer to them as UFOs and non-UFOs (to indicate those detected in the Fe K band with an intermediate ionization/velocity), but they will be correctly considered together in the subsequent correlation analysis in \S4.

The number of sources having papers reporting the detection of WAs is 21/35, therefore the fraction of objects with WAs is at least $\ga$60\%, consistent with previous studies (Reynolds 1997; George et al.~1998; Blustin et al.~2005; McKernan et al.~2007; Winter 2010). If we consider the sources showing absorbers in the form of WAs, UFOs or non-UFOs this increases to 26/35, $\ga$74\%. This suggests that the absolute majority of bright Seyfert 1 galaxies do show some form of ionized X-ray absorption if examined with sufficiently high S/N observations (Winter 2010; Paper I). The fraction of sources showing UFOs is 12/35 ($\ga$34\%) and 8/12 ($\ga$67\%) of these show also WAs. If we consider the UFOs and non-UFOs together we obtain a fraction of 16/35 ($\ga$46\%) and consequently 11/16 ($\ga$69\%) of these sources show also WAs. These fractions might possibly depend also on the inclination of the flow with respect to the line of sight. Considering the fact that some absorbers may have not been detected due to low S/N, variability or simply because some sources have no grating observations or the WAs were not studied in detail, we emphasize that these fractions do represent only lower limits.

\section[]{Warm absorber parameters}

We estimate the lower and upper limits of the distance, mass outflow rate and kinetic power of the WAs following the method outlined in Paper III. An upper limit on the line of sight projected location can be derived from the definition of the ionization parameter reported in \S2 and the requirement that the thickness of the absorber does not exceed its distance to the black hole, $N_\mathrm{H} \simeq n \Delta r < n r$ (e.g. Crenshaw \& Kraemer 2012):

\begin{equation}
r_{\mathrm{max}} \equiv L_{\mathrm{ion}}/\xi N_\mathrm{H},
\end{equation}
the material can not be farther away than this given the observed ionization and column. Instead, an estimate of the minimum distance can be derived from the radius at which the observed velocity corresponds to the escape velocity:

\begin{equation}
 r_{\mathrm{min}} \equiv 2 G M_{\mathrm{BH}}/ v_{\mathrm{out}}^{2}. 
\end{equation}
For the calculation of the mass outflow rate we use the expression derived by Krongold et al.~(2007), which is more appropriate for a biconical wind-like geometry: 

\begin{equation}
\dot{M}_{\mathrm{out}} \equiv f(\delta, \phi) \pi \mu m_\mathrm{p} N_\mathrm{H} v_\mathrm{out} r , 
\end{equation}
where $f(\delta, \phi)$ is a function that depends on the angle between the line of sight to the central source and the accretion disc plane, $\delta$, and the angle formed by the wind with the accretion disc, $\phi$ (see Fig.~12 of Krongold et al.~2007). Instead, $\mu \equiv n_\mathrm{H}/n_\mathrm{e}\simeq 1/1.4$ for Solar abundances. For a roughly vertical disc wind ($\phi$$\simeq$$\pi/2$) and an average line of sight angle of $\delta$$\simeq$$30^{\circ}$ for the Seyfert 1s considered here (Wu \& Han 2001) we have $f(\delta, \phi)$$\simeq$1.5. Full details on the derivation of this formula can be found in the Appendix 2 of Krongold et al.~(2007).

\begin{table*}
\centering
\begin{minipage}{190mm}
\caption{Parameters of the soft X-ray WAs for the type 1 Seyferts in the sample.}
\begin{tabular}{lccccccccc}
\hline
Source         & $\frac{\mathrm{log} M_{BH}}{\mathrm{log} L_{Edd}}$  & Obs & $\mathrm{log} L_{\mathrm{ion}}$ & $\mathrm{log} \xi$ & log$N_{\mathrm{H}}$ & log$v_{\mathrm{out}}$ & $\frac{\mathrm{log} r_{\mathrm{max}}}{\mathrm{log} r_{\mathrm{min}}}$ & $\frac{\mathrm{log} \dot{M}_{\mathrm{out}}^{\mathrm{max}}}{\mathrm{log} \dot{M}_{\mathrm{out}}^{\mathrm{min}}}$ & $\frac{\mathrm{log} \dot{E}_{\mathrm{K}}^{\mathrm{max}}}{\mathrm{log} \dot{E}_{\mathrm{K}}^{\mathrm{min}}}$ \\
               & ($M_{\odot}$/erg~s$^{-1}$)  &     &   (erg~s$^{-1}$)      & (erg~s$^{-1}$~cm) & (cm$^{-2}$) & (cm~s$^{-1}$) & (cm) & (g~s$^{-1}$) & (erg~s$^{-1}$)\\
\hline
NGC~4151$^{\star}$      & $7.1^{30}/45.2$  & C$^1$ & 42.9 & $\sim$2.50    & $\sim$22.50   & $\sim$8.00 & 17.9/17.5 & 25.2/24.8 & 40.9/40.5\\
IC4329A$^{\star}$       & $8.1^{31}/46.2$  & C$^2$,X$^3$ & 44.1 & $-$1.37$\pm$0.06 & 21.12$\pm$0.01 & \dots & 24.4/\dots & \dots/\dots & \dots/\dots\\
                 & &    & & 0.38$\pm$0.07    & 20.94$\pm$0.04    & $\la$7.20 & 22.6/20.1 & 27.7/25.1 & 41.8/39.2\\
                 & &    & & 2.06$\pm$0.05    & 21.49$\pm$0.05    & $\la$7.30 & 20.5/19.9 & 26.1/25.5 & 40.4/39.8\\
NGC~3783$^{\dagger}$       & $7.5^{30}/45.6$  & C$^2$ & 43.5 & 0.40$\pm$0.10    & 21.30$\pm$0.04    & 7.74$\pm$0.02 & 21.8/18.4 & 27.6/24.3 & 42.8/39.4\\
                 &    & & & 2.10$\pm$0.10    & 21.78$\pm$0.07    & 7.70$\pm$0.01 & 19.6/18.5 & 25.9/24.8 & 41.0/39.9\\
                 &    & & & 2.95$\pm$0.07    & 22.00$\pm$0.06   & 7.88$\pm$0.01 & 18.5/18.1 & 25.3/24.9 & 40.9/40.4\\
MCG+8-11-11      & $7.2^{32}/45.3$ & X$^4$  & 43.9  & 2.66$\pm$0.20    & 22.04$\pm$0.24   & \dots & 19.1/\dots & \dots/\dots & \dots/\dots\\
NGC~5548         & $7.8^{30}/45.9$ & C$^2$  & 43.8  & 2.20$\pm$0.20    & 20.78$\pm$0.24    & $\la$7.75 & 20.8/18.7 & 26.1/24.0 & 41.3/39.2\\
NGC~3516$^{\dagger}$        & $7.2^{35}/45.3$ & C$^2$ & 43.7 & 2.40$\pm$0.20    & 21.48$\pm$0.14    & 7.96$\pm$0.04 & 19.8/17.7 & 26.0/24.0 & 41.7/39.6\\
NGC~4593         & $6.7^{30}/44.8$ & C$^2$ & 43.3  & 2.40$\pm$0.20   & 21.30$\pm$0.22    & $\la$7.00 & 19.6/17.1 & 24.7/24.2 & 38.4/37.9\\
Mrk~509$^{\star}$         & $8.1^{30}/46.2$ & C$^5$ & 44.3 & 1.76$\pm$0.14    & 21.31$\pm$0.09    & 7.44$\pm$0.10 & 21.2/19.7 & 26.8/25.2 & 41.4/39.8\\
MCG-6-30-15      & $6.2^{32}/44.3$ & C$^2$ & 43.7 & 0.20$\pm$0.10    & 21.60$\pm$0.11    & \dots & 22.9/\dots & \dots/\dots & \dots/\dots\\
                 &    & & & 2.10$\pm$0.10    & 21.48$\pm$0.14    & \dots & 20.1/\dots & \dots/\dots & \dots/\dots\\
                 &    & & & 3.70$\pm$0.20    & 22.48$^{+0.87}_{-0.29}$   & 8.19$\pm$0.02 & 17.5/16.3 & 25.0/23.7 & 41.1/39.8\\
Ark~120$^{\star}$         & $8.2^{30}/46.3$ & X$^6$ & 44.5 & \dots  & \dots  & \dots & \dots & \dots & \dots\\
Mrk~110          & $7.4^{30}/45.5$ & X$^7$ & 44.3  & \dots  & \dots  & \dots & \dots & \dots & \dots\\
NGC~7469         & $7.1^{30}/45.2$ & X$^8$ & 43.6  & 1.60$^{+0.70}_{-0.40}$    & 20.18$\pm$0.26    & 8.00$\pm$0.22 & 21.7/17.5 & 26.8/22.6 & 42.5/38.3\\
IRAS 05078$+$1626 & $6.9^{33}/45.0$ & X$^9$ & 43.6  & 2.50$^{+1.00}_{-0.40}$    & 24.11$\pm$0.07 & \dots & 17.0/\dots & \dots/\dots & \dots/\dots\\
Mrk~279$^{\dagger}$        & $7.5^{30}/45.6$  & C$^{10}$ & 44.1 & 0.47$\pm$0.07    & 20.09$\pm$0.08    & 7.31$\pm$0.11 & 23.6/19.3 & 27.7/23.4 & 42.0/37.7\\
                 &     & & & 2.49$\pm$0.07    & 20.51$\pm$0.11   & 7.75$\pm$0.10 & 21.1/18.4 & 26.1/23.5 & 41.3/38.7\\
NGC~526A         & $6.2^{33}/44.3$ & & 43.6 & \dots  & \dots  & \dots & \dots & \dots & \dots \\
NGC~3227         & $7.6^{30}/45.7$ & C$^2$,X$^{11}$ & 42.1 & 1.21$\pm$0.10    & 21.04$\pm$0.04    & 7.62$^{+0.27}_{-0.12}$ & 19.9/18.8 & 25.3/24.2 & 40.3/39.1\\
                 &    & & & 2.90$\pm$0.15    & 21.38$^{+0.22}_{-0.13}$   & 8.31$\pm$0.03 & 17.8/17.4 & 24.3/23.9 & 40.7/40.2\\
NGC~7213         & $8.0^{33}/46.1$ & X$^{12}$ & 42.6 & \dots  & \dots  & \dots & \dots & \dots & \dots\\
ESO~511$-$G30    &    &  & 43.7 & \dots  & \dots  & \dots & \dots & \dots & \dots \\
Mrk~79$^{\star}$          & $7.7^{30}/45.8$ & X$^{13}$ & 43.9 & $\sim$1.20    & $\sim$21.00    & \dots & 21.7/\dots & \dots/\dots & \dots/\dots\\
NGC~4051$^{\star}$         & $6.3^{30}/44.4$ & C$^{2, 14}$,X$^{15}$ & 42.2 & $-$0.86$^{+0.09}_{-0.18}$ & 20.49$\pm$0.08   & 7.26$\pm$0.15 & 22.6/18.2 & 27.1/22.8 & 41.3/37.0\\
                 &     & & & 0.60$\pm$0.16    & 20.18$\pm$0.08    & 7.34$\pm$0.06 & 21.4/18.0 & 25.7/22.4 & 40.1/36.8\\
                 &     & & & 1.85$\pm$0.08    & 20.59$\pm$0.09    & 7.79$\pm$0.03 & 19.8/17.1 & 24.9/22.3 & 40.2/37.6\\
                 &     & & & 2.78$\pm$0.17    & 21.28$\pm$0.08    & 7.74$\pm$0.02 & 18.1/17.2 & 24.0/23.1 & 39.1/38.2\\
                 &     & & & 3.35$\pm$0.04    & 22.33$\pm$0.04    & 8.62$\pm$0.01 & 16.5/15.5 & 24.3/23.2 & 41.2/40.2\\
Mrk~766$^{\star}$        & $6.1^{32}/44.2$  & C$^2$ & 43.3 & 2.00$\pm$0.10    & 20.30$^{+0.43}_{-0.22}$    & \dots & 21.0/\dots & \dots/\dots & \dots/\dots\\
                 &     & & & 3.10$\pm$0.20    & 20.78$^{+0.72}_{-0.22}$    & \dots & 19.4/\dots & \dots/\dots & \dots/\dots\\
Mrk~841$^{\star}$         & $7.8^{33}/45.9$  & X$^{16}$ & 43.9 & 1.80$\pm$0.11    & 21.39$\pm$0.13    & $\sim$7.00 & 20.7/20.2 & 25.9/25.4 & 39.6/39.1\\
                 &     & & & 3.10$\pm$0.23    & 22.27$\pm$0.22   & $\sim$8.00 & 18.5/18.2 & 25.6/25.3 & 41.3/41.0\\
Mrk~704          & $7.6^{33}/45.7$ & X$^{17}$ & 43.8  & 1.27$^{+0.27}_{-0.52}$    & 20.30$\pm$0.11    & 8.13$\pm$0.10 & 22.2/17.8 & 27.4/23.0 & 43.4/39.0\\
                 &    & & & 2.70$\pm$0.30    & 20.43$\pm$0.15    & 7.73$\pm$0.08 & 20.6/18.6 & 25.6/23.6 & 40.8/38.8\\
Fairall 9        & $8.4^{30}/46.5$ & C$^{2}$,X$^{18}$ & 44.1 & \dots  & \dots  & \dots & \dots & \dots & \dots\\
ESO~323$-$G77$^{\dagger}$   & $7.4^{33}/45.5$ & X$^{19}$ & 44.0  & \dots  & \dots  & \dots & \dots & \dots & \dots\\
1H~0419$-$577$^{\star}$     & $8.6^{32}/46.7$ & X$^{20}$  & 44.6  & \dots  & \dots  & \dots & \dots & \dots & \dots\\
Mrk~335          & $7.2^{30}/45.3$ & X$^{21}$ & 44.1  & \dots  & \dots  & \dots & \dots & \dots & \dots\\
ESO~198$-$G024   & $8.3^{33}/46.4$ & X$^{22}$ & 44.0  & \dots  & \dots  & \dots & \dots & \dots & \dots\\
Mrk~290$^{\star}$         & $7.7^{33}/45.8$ & C$^{23}$,X$^{23}$ & 43.6 & 2.43$\pm$0.01    & 21.69$\pm$0.03    & 7.65$\pm$0.02 & 19.5/18.8 & 25.7/25.0 & 40.7/40.0\\
Mrk~205$^{\star}$         & $8.6^{34}/46.7$ & X$^{24}$ & 44.2  & \dots  & \dots  & \dots & \dots & \dots & \dots\\
Mrk~590          & $7.7^{30}/45.8$ & C$^{25}$,X$^{25}$ & 43.3 & \dots  & \dots  & \dots & \dots & \dots & \dots \\
H~0557$-$385      & $7.6^{33}/45.7$ & X$^{26}$  & 43.4 & 0.50$\pm$0.18    & 21.30$\pm$0.13    & \dots & 21.6/\dots & \dots/\dots & \dots/\dots\\
                 &    & & & 2.33$\pm$0.03    & 22.11$\pm$0.03   & \dots & 19.0/\dots & \dots/\dots & \dots/\dots\\
TON~S180         & $7.1^{32}/45.2$ & C$^{27}$  & 44.2 & \dots  & \dots  & \dots & \dots & \dots & \dots\\
PG~1211$+$143$^{\star}$   & $8.2^{30}/46.3$ & X$^{28}$ & 44.3  & \dots    & \dots  & \dots & \dots & \dots & \dots\\
Ark~564          & $6.1^{32}/44.2$ & C$^{2}$,X$^{29}$ & 44.6 & $-$0.86$\pm$0.10 & 19.95$\pm$0.09 & $\la$7.00 & 25.5/18.5 & 29.3/22.3 & 43.0/36.0 \\
                 &    & &  & 0.87$\pm$0.07 & 20.38$\pm$0.04    & $\la$7.00  & 23.3/18.5 & 27.5/22.7 & 41.2/36.4\\
                 &    & &  & 2.58$\pm$0.05 & 20.54$\pm$0.17    & $\la$7.18  & 21.5/18.2 & 26.0/22.7 & 40.1/36.7\\
\hline
\end{tabular}
\scriptsize{{\em Notes.} C and X stand for grating observations with \emph{Chandra} or \emph{XMM-Newton}, respectively.  The $^{\star}$ and $^{\dagger}$ mark the sources with detected Fe K absorbers identified with UFOs and non-UFOs in Paper II, respectively. $^1$ Kraemer et al.~(2005); $^2$ McKernan et al.~(2007); $^3$ Steenbrugge et al.~(2005); $^4$ Matt et al.~(2006); $^5$ Yaqoob et al.~(2003); $^6$ Vaughan et al.~(2004); $^7$ Cardaci et al.~(2011); $^8$ Blustin et al.~(2003); $^9$ Svoboda et al.~(2010); $^{10}$ Costantini et al.~(2007); $^{11}$ Markowitz et al.~(2009); $^{12}$ Starling et al.~(2005); $^{13}$ Gallo et al.~(2011); $^{14}$ Lobban et al.~(2011); $^{15}$ Pounds \& Vaughan (2011); $^{16}$ Longinotti et al.~(2010); $^{17}$ Laha et al.~(2011); $^{18}$ Emmanoulopoulos et al.~(2011) $^{19}$  Jim{\'e}nez-Bail{\'o}n et al.~(2008); $^{20}$ Pounds et al.~(2004); $^{21}$ Gondoin et al.~(2002); $^{22}$ Porquet et al.~(2004); $^{23}$ Zhang et al.~(2011); $^{24}$ Reeves et al.~(2001); $^{25}$ Longinotti et al.~(2007); $^{26}$ Ashton et al.~(2006); $^{27}$  R{\'o}{\.z}a{\'n}ska et al.~(2004); $^{28}$ Pounds et al.~(2003); $^{29}$ Smith et al.~(2008); $^{30}$ Peterson et al.~(2004); $^{31}$ Markowitz~(2009); $^{32}$ Bian \& Zhao (2003a); $^{33}$ Wang \& Zhang (2007); $^{34}$ Wandel \& Mushotzky (1986); $^{35}$ Onken et al.~(2003).}       
\end{minipage}
\end{table*}

This expression for the mass outflow rate has also the important advantage of not relying on the estimate of the covering and filling factors. This is due to the fact that it takes into account only the net observed thickness of the gas, allowing for clumping in the flow. Thus, there is not the need to include a linear (or volume) filling factor, since we are interested in estimating the net flow of mass, starting from the observed column density and velocity. Moreover, the covering factor is implicitly taken into account by the function $f(\delta, \phi)$ when	calculating the	area filled by the gas, constrained between the inner and outer conical surfaces. The assumptions are that the thickness of the wind between the two conical surfaces is constant with $\delta$ and that this is smaller than the distance to the source. However, as already noted in Paper III, we obtain equivalent results including a clumpiness factor of $\sim \Delta r/r$ along the line of sight in the spherical approximation case and using a covering fraction $C_\mathrm{f} \simeq 0.2 f(\delta, \phi) \simeq 0.4$, which is consistent with observations (e.g., Blustin et al.~2005; McKernan et al.~2007). Moreover, this expression has the same parameter dependencies as that recently employed by Crenshaw \& Kraemer (2012). Equation (3) is actually more conservative, yielding mass outflow rates that are roughly a factor of two lower.   

Neglecting additional acceleration of the outflow, i.e. assuming that it has reached a constant terminal velocity, the kinetic (or mechanical) power can consequently be derived as: 

\begin{equation}
\dot{E}_\mathrm{K} \equiv \frac{1}{2} \dot{M}_{\mathrm{out}} v_\mathrm{out}^2. 
\end{equation}
The estimates of these parameters are reported in Table~1. We also calculated the outflow momentum rate as $\dot{P}_\mathrm{out} \equiv \dot{M}_\mathrm{out} v_\mathrm{out}$ and subsequently compared it to the momentum flux of the radiation field, $\dot{P}_\mathrm{rad} \equiv L_\mathrm{bol}/c$. The bolometric luminosity $L_\mathrm{bol}$ is estimated as $L_\mathrm{bol} = k_\mathrm{bol} L_\mathrm{ion}$, where $k_\mathrm{bol}$$\simeq$10 (McKernan et al.~2007; Vasudevan \& Fabian 2009; Lusso et al.~2010).

  \begin{figure}
  \centering
   \includegraphics[width=7.1cm,height=7.1cm,angle=0]{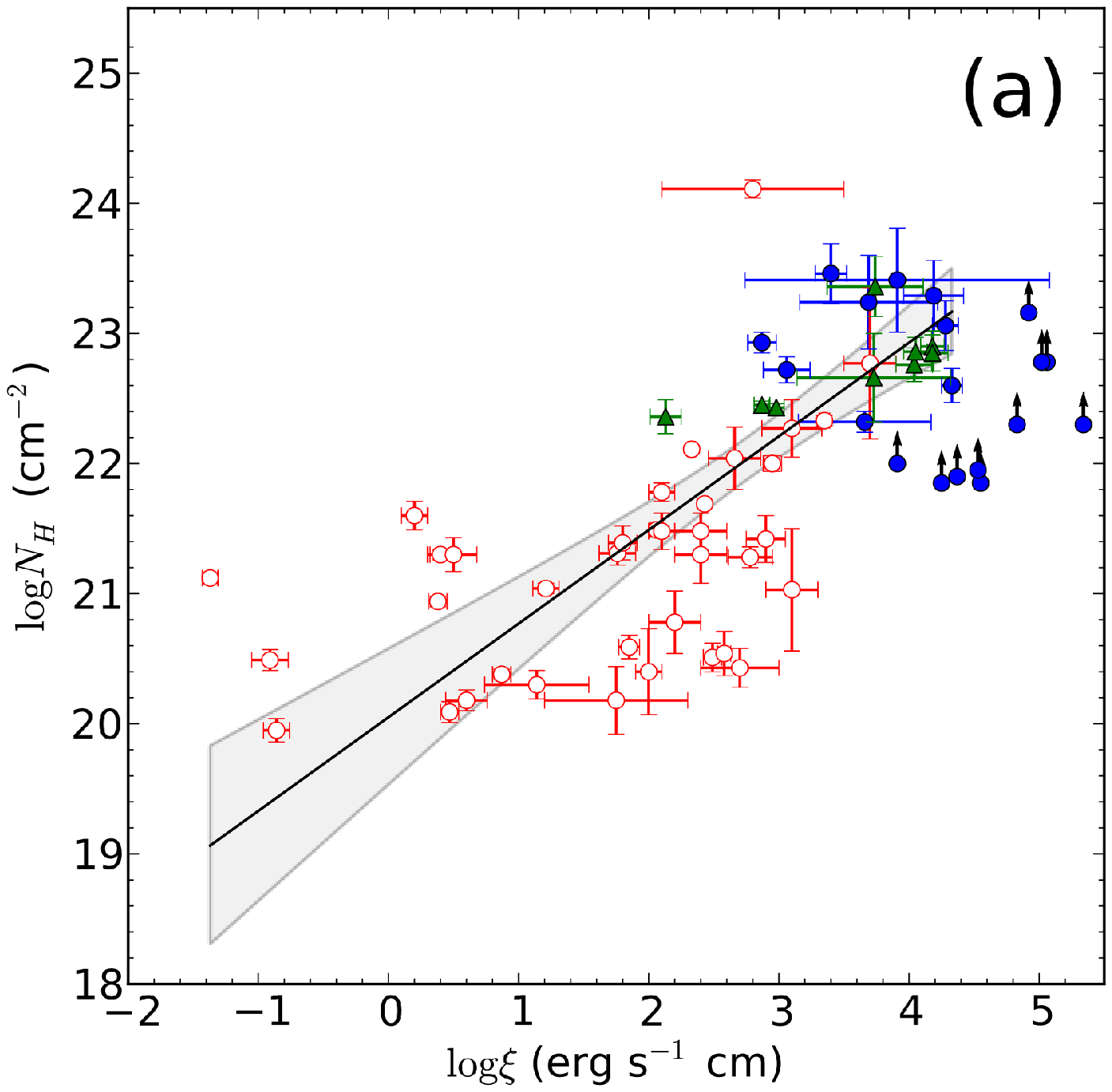}
   \includegraphics[width=7.1cm,height=7.1cm,angle=0]{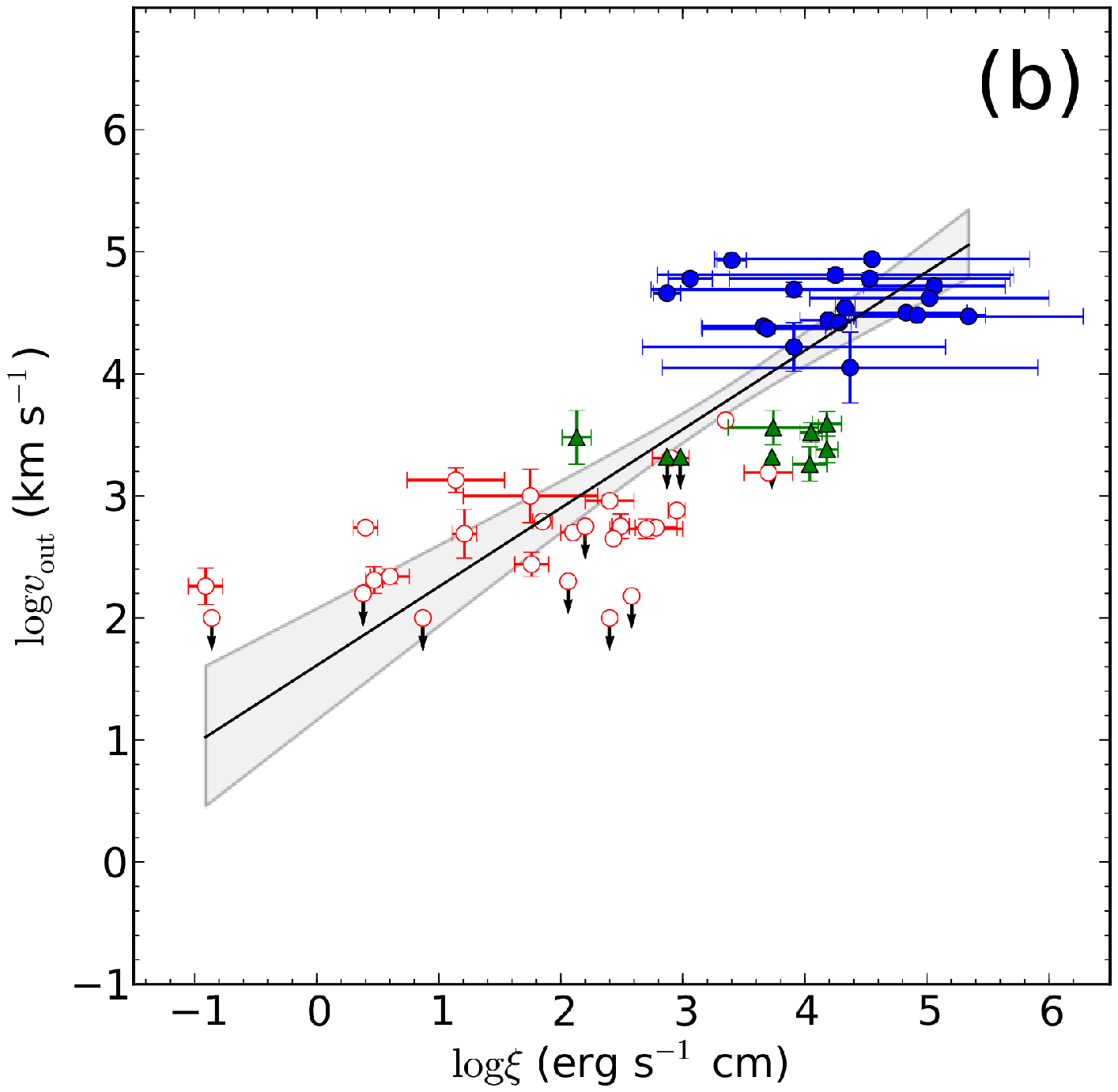}
   \includegraphics[width=7.25cm,height=7.1cm,angle=0]{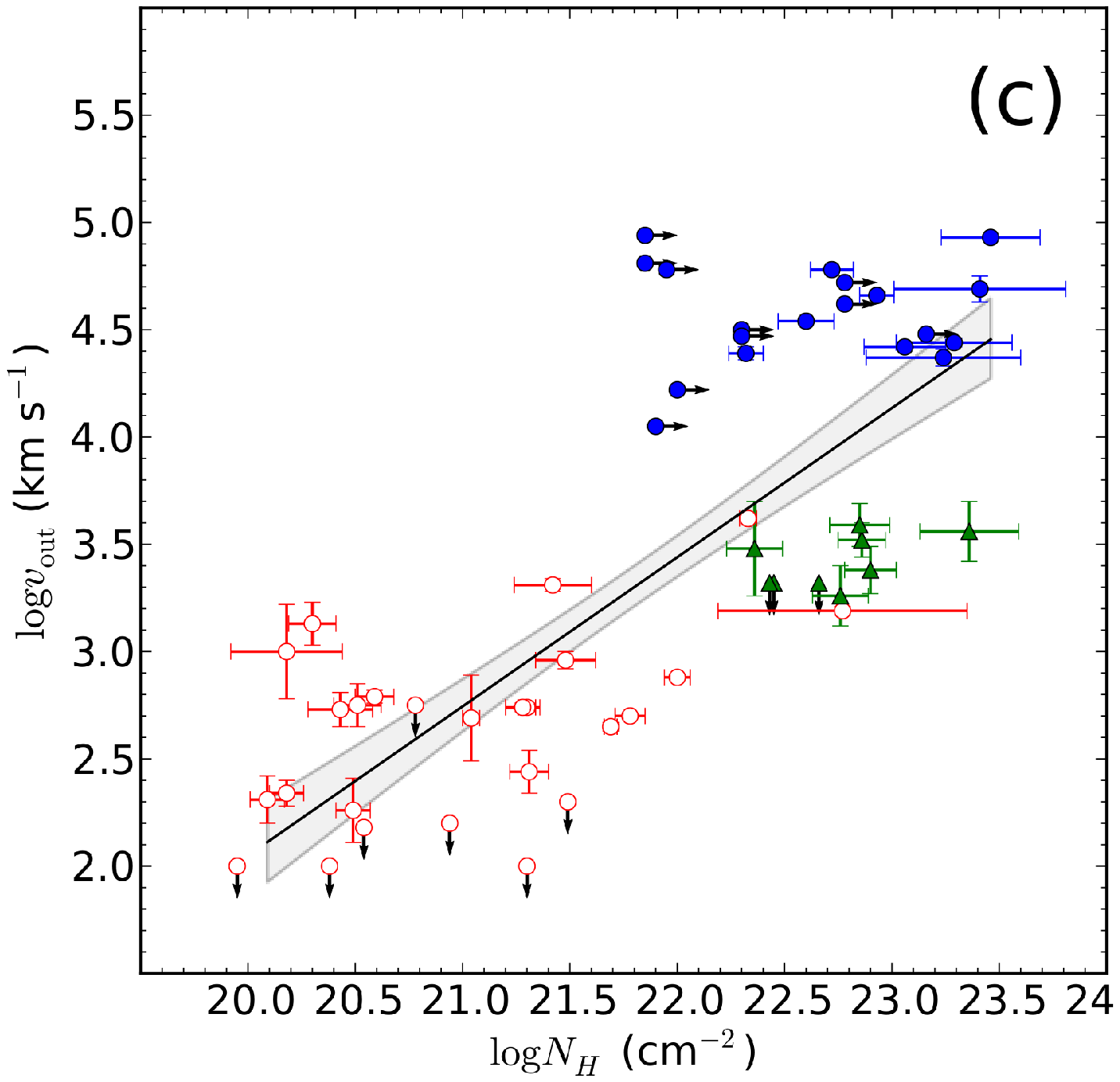}
   \caption{Correlations for measured outflow parameters. Scatter plots of log$\xi$ vs. logN$_\mathrm{H}$ (panel a), log$\xi$ vs. log$v_\mathrm{out}$ (panel b) and logN$_\mathrm{H}$ vs. log$v_\mathrm{out}$ (panel c) for the WAs (red), non-UFOs (green) and UFOs (blue). The arrows indicate the lower or upper limits. The solid lines represent the best-fit linear correlations and the gray shadowed areas indicate the 2$\sigma$ confidence bands.}
    \end{figure}

\section[]{Correlation analysis}

\subsection[]{Measured absorber parameters}

Initially, we consider the data from all WAs, UFOs and non-UFOs together. We compare their ionization parameter, velocity and column density and search for possible correlations. The large dynamic range of $\sim$6 orders of magnitude in ionization, $\sim$4 in column density and $\sim$3 in outflow velocity allows, for the first time, to investigate the general relations and trends among these parameters.
We fit the power-law model $\log y = a \log x +b$ to the datasets using the BCES (bivariate correlated errors and intrinsic scatter) regression method (Akritas \& Bershady 1996) which takes into account measurement errors in both the ``$X$'' and ``$Y$'' coordinates as well as the intrinsic scatter. This method has been widely used in fitting datasets in the astronomical community (e.g., Sani et al.~2011). No a priori dependent variable in the fitting is assumed and we treat the variables symmetrically. Uncertainties on the parameters derived from the fits are estimated after carrying out 10,000 bootstrap resamples of the data. In order to estimate the significance of these correlations we calculated the Pearson coefficient $R_\mathrm{P}$.

The results of the fits are listed in Table~2 and the scatter plots and best-fit relations are shown in Fig.~1\footnote{In calculating the correlations we used only the points constrained within their errors of measure. Instead, in the figures we plot also those with upper or lower limits for completeness.}.  All three correlations have a high statistical significance($\ge 6 \sigma$). We note that the wide dynamic range covered by the observed absorption components allows us to fill the whole parameter space, with the WAs and the UFOs at the two sides.

\begin{table*}
\centering
\begin{minipage}{155mm}
\caption{Results of the linear regression and partial correlation analysis for the measured absorbers parameters.}
\begin{tabular}{c@{\hspace{0.3cm}}c@{\hspace{0.3cm}}c@{\hspace{0.3cm}}c@{\hspace{0.3cm}}c@{\hspace{0.3cm}}c@{\hspace{0.3cm}}c@{\hspace{0.3cm}}c@{\hspace{0.3cm}}c@{\hspace{0.3cm}}c@{\hspace{0.3cm}}c@{\hspace{0.3cm}}c@{\hspace{0.3cm}}c@{\hspace{0.3cm}}c}
\hline
x & y & $a$ & dev($a$)  & $b$ & dev($b$) & scatt & $R_\mathrm{P}$ & dof & $P_\mathrm{null}$ & z & $\tau_\mathrm{K}$ & $\sigma_\mathrm{K}$ & $P_\mathrm{part}$\\
\hline
\multicolumn{13}{c}{All absorbers together}\\
\hline
log$\xi$ & $\mathrm{logN}_\mathrm{H}$ & 0.72 & 0.12 & 20.00 & 0.33 & 0.71 & 0.73 & 55 & $1.0\times10^{-10}$ & log$v_\mathrm{out}$ & 0.5 & 0.09 & $1.0\times 10^{-10}$\\
log$\xi$ & log$v_\mathrm{out}$ & 0.65 & 0.10 & 6.61 & 0.35 & 0.52 & 0.79 & 45 & $4.0\times 10^{-11}$ & logN$_\mathrm{H}$ & 0.3 & 0.07 & $1.0\times 10^{-4}$\\
logN$_\mathrm{H}$ & log$v_\mathrm{out}$ & 0.69 & 0.09 & $-$6.84 & 1.97 & 0.45 & 0.77 & 65 & $3.0\times 10^{-14}$ & log$\xi$ & 0.3 & 0.09 & $5.0\times 10^{-4}$\\
\hline
\multicolumn{13}{c}{WAs}\\
\hline
log$\xi$ & $\mathrm{logN}_\mathrm{H}$ & 0.73 & 0.13 & 19.92 & 0.29 & 0.83 & 0.50 & & $1.5\times 10^{-3}$ & & & &\\
log$\xi$ & log$v_\mathrm{out}$ & 0.31 & 0.08 & 7.19 & 0.18 & 0.30 & 0.69 & & $1.0\times 10^{-3}$ & & & & \\
logN$_\mathrm{H}$ & log$v_\mathrm{out}$ & 0.48 & 0.21 & $-$2.46 & 4.47 & 0.30 & 0.52 & & $2.0\times 10^{-2}$ & & & &\\
\hline
\multicolumn{13}{c}{UFOs and non-UFOs}\\
\hline
log$\xi$ & $\mathrm{logN}_\mathrm{H}$ & 0.62 & 0.53 & 20.62 & 1.98 & 0.27 & 0.41 & & $9.0\times 10^{-2}$ & & & &\\
log$\xi$ & log$v_\mathrm{out}$ & 0.63 & 1.09 & 6.72 & 4.46 & 0.60 & 0.23 & & $2.6\times 10^{-1}$ & & & &\\
logN$_\mathrm{H}$ & log$v_\mathrm{out}$ & 1.43 & 1.30 & $-$23.74 & 30.10 & 0.79 & 0.29 & & $2.9\times 10^{-1}$ & & & &\\
\hline
\end{tabular}
\scriptsize{{\em Notes.} $a$ and $b$ are the slope and intercept of the linear correlation fits with standard deviations dev($a$) and dev($b$), respectively. scatt represents the internal scatter in units of dex. $R_\mathrm{P}$ is the Pearson coefficient. dof is the number of degrees of freedom. $P_\mathrm{null}$ is the probability of the null hypothesis that there is no correlation between x and y. z is the third variable against which the partial correlation analysis is performed. $\tau_\mathrm{K}$ and $\sigma_\mathrm{K}$ are the Kendall's partial correlation coefficient and the variance, respectively. $P_\mathrm{part}$ is the null hypothesis probability of the partial correlation.}       
\end{minipage}
\end{table*}

In particular, the slope $0.72\pm0.12$ of the correlation between the ionization parameter and column density in Fig.~1a suggests that the column density is higher for more ionized absorbers\footnote{In Fig.~1a we note a possible WA outlier with high column density, $N_\mathrm{H} \sim 10^{24}$~cm$^{-2}$, and a relatively high ionization, $\mathrm{log}\xi \sim$2.5~erg~s$^{-1}$~cm. As reported in Table~1, this corresponds to the Seyfert 1.5 galaxy IRAS~05078$+$1626. However, given that it has not a velocity estimate, this is not considered anymore in all the successive relations.}. From Fig.~1b we observe a linear relation with slope $0.65\pm0.10$ between the ionization and velocity, indicating that the faster outflows are also more ionized. Finally, Fig.~1c shows a relation between column density and velocity with slope $0.69\pm0.09$, which indicates that the faster outflows have also higher columns. In general, these correlations suggest that the faster the outflow, the higher the column density and ionization parameter. The possible systematics and selection effects affecting these relations are discussed later in \S4.3.

In order to test whether the correlations between two parameters are driven by their mutual dependence on a third one, we then performed also a partial correlation analysis (Akritas \& Siebert 1996). We quantify this effect using the Kendall's partial correlation coefficient, $\tau_\mathrm{K}$, which takes into account also censored data\footnote{The partial correlation analysis takes into account only the points constrained within their errors of measure and those with upper limits.}. The results of the partial correlation analysis are reported in the last three columns of Table~2. This test indicates that the correlations between the different parameters are only marginally interdependent. 

We performed also a correlation analysis considering separately the WAs and the other absorbers (UFOs and non-UFOs). As already noted in \S2, the distinction between UFOs and non-UFOs is only arbitrary and based on their velocity being higher or lower than 10,000~km/s. In fact, their parameter distributions reported in Paper II and Paper III are not bimodal. This test was done in order to check if there are significant differences when considering the two populations separately. The results are reported in Table~2. 
Due to the large error bars, smaller number of data points and reduced dynamic range of the two separated populations, their parameter values are less constrained and the significance is lower, especially for the highly ionized absorbers. 
However, a possible difference can be noted in the best-fit value of the slope of the relations between log$\xi$--log$v_\mathrm{out}$ and log$N_\mathrm{H}$--log$v_\mathrm{out}$, the one of the highly ionized absorbers being steeper, although they are still consistent given the large uncertainties. This point will be addressed later in the discussion section. Given the limited quality of the current datasets we are not in a position to strongly constrain possible differences between the separated correlations. 

Finally, we note that Blustin et al.~(2005) attempted a similar correlation analysis of the WAs collecting their parameters from the literature. They considered a sample of 23 Seyferts and quasars, finding useful information for only 14 of them. They found a significant correlation between log$\xi$ and log$N_\mathrm{H}$, as reported also by other authors (Holczer et al.~2007; Behar 2009). Excluding instead the few fast and highly ionized outflows reported at that time, they found only a very marginal correlation between log$\xi$ and log$v_\mathrm{out}$ and no significant correlation between log$v_\mathrm{out}$ and log$N_\mathrm{H}$. However, it is important to bear in mind that the inability to find a significant correlation does not necessarily exclude its existence. In fact, after more than seven years of additional observations and analysis, we can now define a larger sample of sources (35) with higher S/N observations, which allowed us to increase the number and quality of the reported WAs components and, thus, better constrain the correlations between the parameters. The inclusion of the UFOs also allows to significanly expand the parameter space.

  \begin{figure*}
  \centering
   \includegraphics[width=5.5cm,height=8cm,angle=-90]{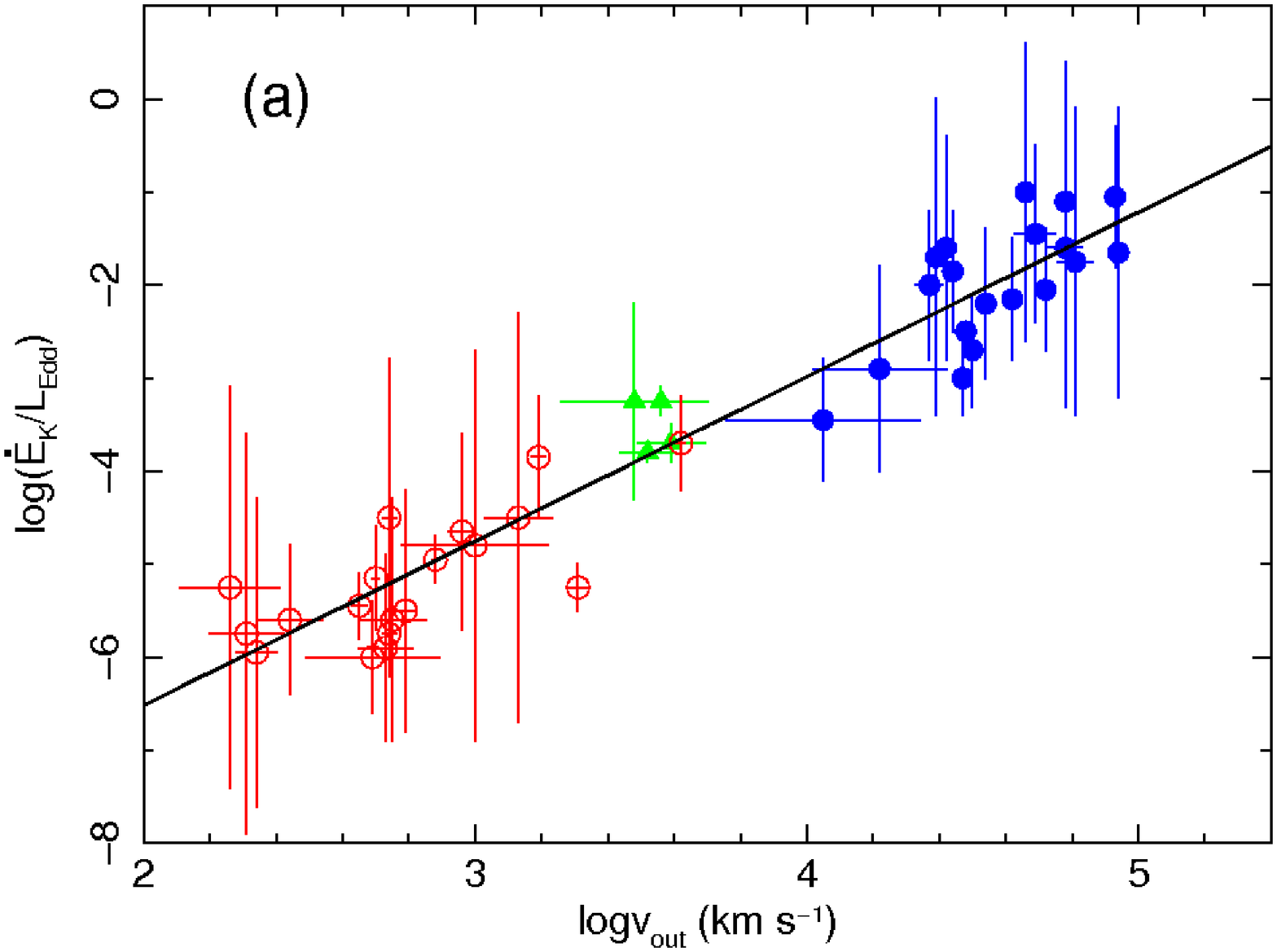}
   \includegraphics[width=5.5cm,height=8cm,angle=-90]{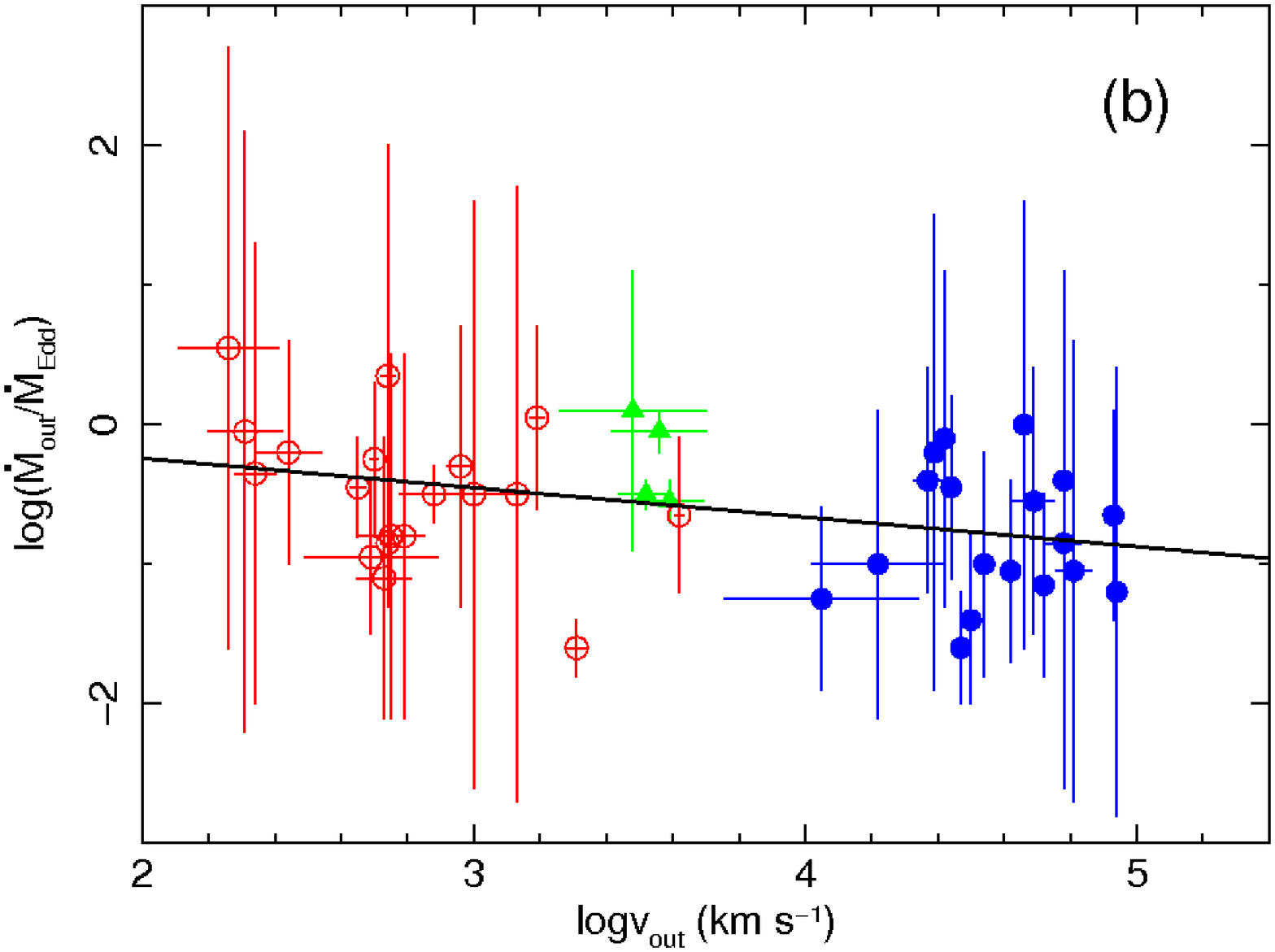}
   \caption{Comparison between the outflow velocity log$v_\mathrm{out}$, the outflow kinetic power log$(\dot{E}_\mathrm{K}/L_\mathrm{Edd})$ (panel a) and the mass outflow rate log$(\dot{M}_\mathrm{out} / \dot{M}_\mathrm{Edd})$ (panel b), respectively. The points relative to the WAs (red open circles), non-UFOs (green filled triangles) and UFOs (blue filled circles) are reported. The error bars indicate the upper and lower limits and the points are the average between the two. The solid lines indicate the best-fit linear regression curves.}
    \end{figure*}

\subsection[]{Derived outflow parameters}

We checked for possible correlations/trends among the derived outflow parameters of the WAs listed in Table~1 and those of the UFOs and non-UFOs reported in Paper III. Given that we can only estimate upper and lower limits, in order to carry out the fits we performed 10,000 Monte Carlo simulations for each dataset, considering a random value between the lower and upper limits for each point (assuming a uniform distribution in log space). We then calculated the linear regression and the corresponding statistics. 

The average black hole mass of the sources in the sample is log$\left(M_{BH}/M_\odot\right)$$\simeq$7.4 with a standard deviation of $\simeq$0.7~dex. The difference is not large, but in order to take into account the expected scaling of the outflow parameters with mass, we normalized the distance to the Schwarzschild radius $r_\mathrm{s} \equiv 2GM_\mathrm{BH}/c^2$, the kinetic power to the Eddington luminosity $L_\mathrm{Edd} \equiv 4\pi G M_\mathrm{BH} m_p c/\sigma_T \simeq 1.26\times 10^{38}\left( M_\mathrm{BH}/M_\odot \right)$ erg~s$^{-1}$, the momentum to the Eddington momentum rate $P_\mathrm{Edd} \equiv L_\mathrm{Edd}/c$ and the mass outflow rate to the Eddington rate $\dot{M}_\mathrm{Edd} \equiv \eta L_\mathrm{Edd}/c^2$, where $\eta \simeq 0.1$.

The parameters and significance of the correlations are listed in Table~3. The plots are reported in Fig.~2--4\footnote{For each of these relations we considered only the points constrained between the errors or upper/lower limits in both the X and Y axes. Therefore, the number of points can slightly differ from one plot to the other.}. These show that, even if the difference between the lower and upper limits can be large in some cases, the wide dynamic range of the parameters still allows us to estimate significant correlations/trends among them. The plots of the velocity with respect to the kinetic power and mass outflow rate are reported in Fig.~2. In Fig.~2a we note a strong trend of increasing the kinetic power for increasing velocity, with a positive slope of $1.77$$\pm$$0.14$. This is close to the trend expected for a roughly constant mass outflow rate. In fact, from Fig.~2b we note that, besides the large uncertainties, $\dot{M}_\mathrm{out}$ does not vary much with $v_\mathrm{out}$, although there is a possible weak trend of decreasing mass outflow rate with increasing velocity with slope $-0.21$$\pm$$0.14$ (see Table~3).

Instead, in Fig.~3 we show the plots of the different parameters with respect to the line of sight projected distance in units of $r_\mathrm{s}$. From panels a, b and c we note very significant trends of decreasing the ionization parameter, column density and velocity with distance, with slopes of $-0.58$$\pm$$0.04$, $-0.44$$\pm$$0.04$ and $-0.40$$\pm$$0.03$, respectively (see Table~3). In particular, we checked that the correlation between the velocity and distance is not an induced relation from the fact that the lower limits have been estimated using equation (2) in \S3, i.e. assuming that the observed velocity is equivalent to the escape velocity, as this relation is independently confirmed using the upper limits alone derived from equation (1), as discussed also later in \S4.3. In panel d and e we can observe a weak increase/decrease of the mass outflow rate/momentum rate for increasing distance, with slopes of $0.16$$\pm$$0.07$ and $-0.22$$\pm$$0.07$, respectively. Instead, in panel f we can note a more pronounced and significant trend of increasing the observed outflow mechanical power with decreasing distance, going from the WAs to the UFOs, with a slope of $-0.60$$\pm$$0.09$. 

It should be noted that the whole parameter space is essentially uniformly filled, with distances ranging from $\sim$100~$r_\mathrm{s}$ from the black hole up to $\sim$kpc scales. The UFOs occupy the lower end of the distribution at the smaller distances, where the ionization, column and velocity are higher. In particular, extrapolating the relations reported in Table~3 and Fig.~3 to the innermost possible radii of the order of log($r/r_\mathrm{s})$$\simeq$1 we obtain that the ionization of the gas reaches very high values of log$\xi$$\simeq$5--6~erg~s$^{-1}$~cm, the column becomes mildly Compton-thick, $N_\mathrm{H}$$\simeq$$10^{24}$~cm$^{-2}$, and the outflow velocity approaches significant fractions of the speed of light. The values of the parameters gradually decrease with increasing distance, going from the UFOs to the WAs.

Finally, Fig.~4 shows a significant (99\% confidence level) linear relation between the radiation momentum rate, $\dot{P}_\mathrm{rad}$$=$$L_\mathrm{bol}/c$, and the outflow momentum rate of the UFOs, $\dot{P}_\mathrm{out}$, with a slope of $0.76$$\pm$$0.19$. The possible physical implications of this and the previous relations will be briefly addressed in discussion section \S5.

\begin{table*}
\centering
\begin{minipage}{120mm}
\caption{Linear regression results for the derived outflow parameters.}
\begin{tabular}{cccccccc}
\hline
x & y & $a$ & dev($a$)  & $b$ & dev($b$) & $R_\mathrm{P}$ & $P_\mathrm{null}$\\
\hline
log$v_\mathrm{out}$ & log$(\dot{E}_\mathrm{K}/L_\mathrm{Edd})$ & 1.77 & 0.14 & $-$18.89 & 1.19 & 0.89 & $3.8\times 10^{-15}$\\[4pt]
log$v_\mathrm{out}$ & log$(\dot{M}_\mathrm{out}/\dot{M}_\mathrm{Edd})$ & $-$0.21 & 0.14 & 1.24 & 1.19 & $-$0.22 & $1.6\times 10^{-1}$\\
\hline
log$\left(r/r_\mathrm{s}\right)$ & log$\xi$ & $-$0.58 & 0.04 & 5.80 & 0.21 & $-$0.85 & $8.0\times 10^{-16}$\\[4pt]
log$\left(r/r_\mathrm{s}\right)$ & log$N_\mathrm{H}$ & $-$0.44 & 0.04 & 24.21 & 0.20 & $-$0.86 & $1.3\times 10^{-13}$\\[4pt]
log$\left(r/r_\mathrm{s}\right)$ & log$v_\mathrm{out}$ & $-$0.40 & 0.03 & 10.47 & 0.11 & $-$0.89 & $1.6\times 10^{-15}$\\[4pt]
log$\left(r/r_\mathrm{s}\right)$ & log$(\dot{M}_\mathrm{out}/\dot{M}_\mathrm{Edd})$ & 0.16 & 0.07 & $-$1.29 & 0.32 & 0.38 & $5.0\times 10^{-3}$\\[4pt]
log$\left(r/r_\mathrm{s}\right)$ & log$(\dot{P}_\mathrm{out}/\dot{P}_\mathrm{Edd})$ & $-$0.22 & 0.07 & $-$0.44 & 0.33 & $-$0.45 & $7.2\times 10^{-4}$\\[4pt]
log$\left(r/r_\mathrm{s}\right)$ & log$(\dot{E}_\mathrm{K}/L_\mathrm{Edd})$ & $-$0.60 & 0.09 & $-$0.89 & 0.38 & $-$0.73 & $3.1\times 10^{-10}$\\
\hline
log$\dot{P}_\mathrm{rad}$ & log$\dot{P}_\mathrm{out}$ & 0.76 & 0.19 & 8.68 & 6.41 & 0.56 & $1.0\times 10^{-2}$\\
\hline
\end{tabular}
 \scriptsize{{\em Notes.} $a$ and $b$ are the slope and intercept of the linear correlation fits with standard deviations dev($a$) and dev($b$), respectively. $R_\mathrm{P}$ is the Pearson coefficient. $P_\mathrm{null}$ is the probability of the null hypothesis.}
\end{minipage}
\end{table*}

\subsection[]{Possible systematics and selection effects}

In the calculation of the correlations in the previous sections we took into account the uncertainties in the ionization parameter, column density and outflow velocity as reported in Table~2. However, when considering such a large dataset and especially when collecting results from the literature, it is important to bear in mind the existence of possible systematics and selection effects. 

As already discussed in Paper II, different assumptions of the velocity broadening of the lines can generate some variations in the estimated column density. Regarding the UFOs, in Paper II we already took into account this effect testing different velocity broadenings for the lines that were not resolved and included the larger error bars. For the column densities of the WAs collected from the literature, this effect is marginal given that the estimates were derived using high energy resolution data. Another parameter that can affect the estimate of the column density is the assumed elemental abundance. However, as already discussed in Paper III, even allowing for a factor of two difference with respect to the standard Solar values, the discrepancy in the column density is $\la$0.2~dex, within the typical errors of measure. 
The fact that the column densities do not exceed the value of $N_\mathrm{H}$$\simeq$$10^{24}$~cm$^{-2}$, especially noticeable in Fig.~1c, could in principle be affected by the fact that the photo-ionization code \emph{Xstar} can not treat Compton-thick absorbers.
However, the data do not seem to require significantly higher columns as a good spectral modelling of the highly ionized UFOs in Paper II was already obtained with columns in the range $N_\mathrm{H}$$=$$10^{22}$--$10^{24}$~cm$^{-2}$. 

It is known that the ionization parameter log$\xi$ has a dependence on the assumed ionizing continuum. In Paper III we estimated that the possible uncertainty on the continuum slope may induce a maximum systematic error of $\sim$0.4~dex on the ionization parameter, within the typical internal scatter of the relations shown in Fig.~1 (panels a and b). Moreover, the large range in observed ionization states reduces the importance of this effect when performing the linear regression fits.

Considering the column density and ionization in Fig.~1a, there might be some possible selection effects at the two ends of the distribution. In fact, the limited S/N of the observations could have hampered the detection of the weak spectral features from absorbers with high ionization and low column density. On the other hand, the fact that we can observe WAs only for type 1 sources could have limited the relevance of lowly ionized/neutral absorbers with high column densities. However, these latter types of absorbers, usually found only in type 2 sources, have intrinsic velocity consistent with zero and probably have a different origin than the AGN outflows studied here, such as the $\sim$pc scale molecular torus or large scale dust lanes in the host galaxy itself. The combination of these two possible selection effects could have induced a slight steepening of the log$\xi$--log$N_\mathrm{H}$ relation and of the radial density profile successively discussed in \S5.2. Moreover, these estimates do not take into account the possible presence of additional fully ionized material, which does not imprint any observable spectral absorption features. 

We note that another important parameter of these outflows is their inclination with respect to the line of sight. Unfortunately, the estimate of this parameter is not well constrained for each source yet, but the typical inclination of type 1 Seyferts is $\simeq$30$^{\circ}$, with a range between $\sim$10$^{\circ}$--60$^{\circ}$ (Wu \& Han 2001). Therefore, this ensures that the difference among the sources of our sample is not large. However, this might contribute to some of the internal scatter observed in Fig.~1. Anyway, given the large dynamic range, small differences between sources do not significantly affect the derived scale relations.

  \begin{figure*}
  \centering
   \includegraphics[width=5.6cm,height=8cm,angle=-90]{fig_3a.ps}
   \includegraphics[width=5.6cm,height=8cm,angle=-90]{fig_3b.ps}
   \includegraphics[width=5.6cm,height=8cm,angle=-90]{fig_3c.ps}
   \includegraphics[width=5.6cm,height=8cm,angle=-90]{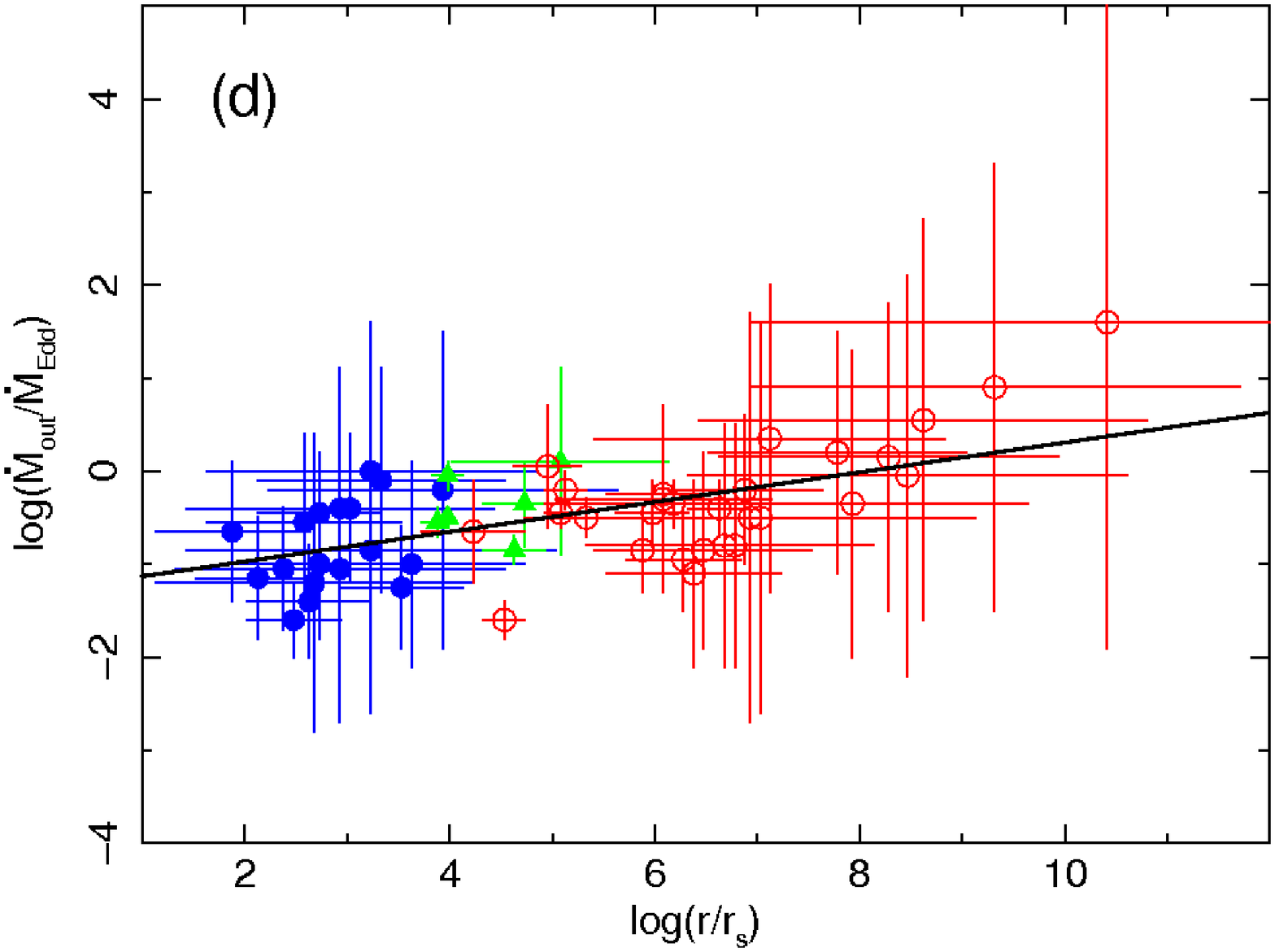}
   \includegraphics[width=5.6cm,height=8cm,angle=-90]{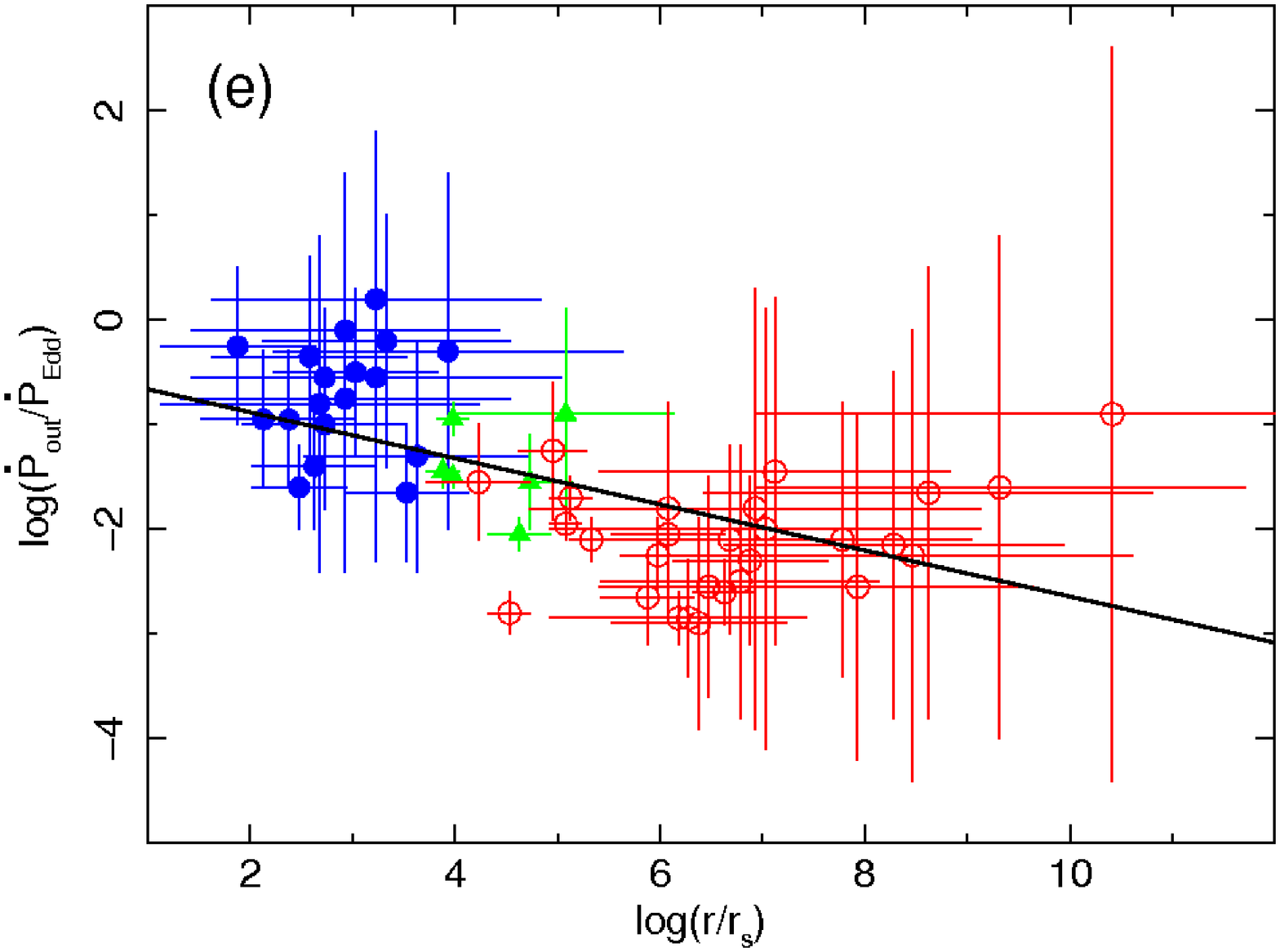}
   \includegraphics[width=5.6cm,height=8cm,angle=-90]{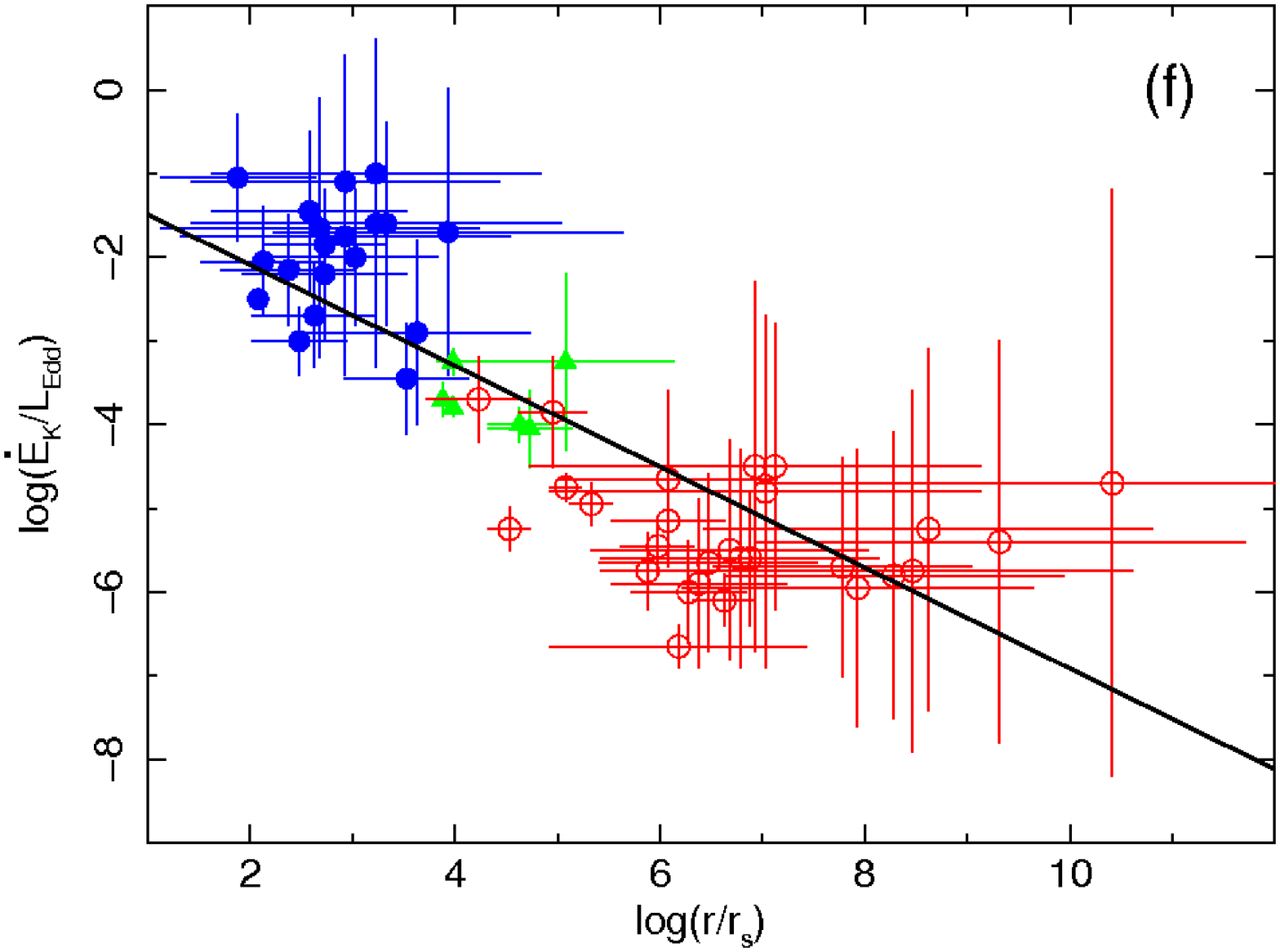}
   \caption{Comparison between the line of sight projected distance log$(r/r_\mathrm{s})$ (in units of the Schwarzschild radius) and the ionization parameter log$\xi$ (panel a), the column density log$N_\mathrm{H}$ (panel b), the outflow velocity log$v_\mathrm{out}$ (panel c), the mass outflow rate log$(\dot{M}_\mathrm{out}/\dot{M}_\mathrm{Edd})$ (panel d), the outflow momentum rate log$(\dot{P}_\mathrm{out}/\dot{P}_\mathrm{Edd})$ (panel e) and the outflow kinetic power log$(\dot{E}_\mathrm{K}/L_\mathrm{Edd})$ (panel f), respectively. The points relative to the WAs (red open circles), non-UFOs (green filled triangles) and UFOs (blue filled circles) are reported. The solid lines indicate the best-fit linear regression curves. Assuming the typical black hole mass of $M_\mathrm{BH} \simeq 10^7 M_{\odot}$, the distance scale can be easily converted from Schwarzschild radii to parsec considering that $1 \mathrm{pc} \simeq 10^{6} r_\mathrm{s}$.}
    \end{figure*}

As already discussed in Paper III, the expression for the mass outflow rate used in \S3 introduces a possible systematic source of uncertainty from the assumed typical inclination and opening angle of the wind, which however is constrained within a factor of $\sim$0.3~dex (Krongold et al.~2007). Instead, the typical systematic uncertainty on the black hole masses derived using the reverberation mapping technique is $\la$0.5~dex (Peterson et al.~2004). 

Regarding the outflow velocity of the absorbers, as already discussed in Paper II, we might be losing some of the components with the highest blue-shift due to the fact that they are usually detected in the Fe K band at E$>$7~keV where both the energy resolution and effective area of the EPIC-pn instrument on board \emph{XMM-Newton} are worse and the detector has essentially no sensitivity above E$\simeq$10~keV.  

We also checked that the methods used to estimate the the lower and upper limits of the distance of the absorbers reported in equations (1) and (2) do not introduce significant systematics in the relations reported in Fig.~2 and 3. For instance, the same dependence between the velocity and distance with a slope consistent with $\simeq$$-$0.4 shown in Fig.~3c is independently found also when considering only the upper limits of the distance derived from the ionization parameter in equation (1). 

All these sources of uncertainties can contribute to the significant internal scatter of $\ga$0.5~dex observable in the plots in Fig.~1. Moreover, we should bare in mind also the possibility that inhomogeneities and variability of the absorbers and also the distinct analysis methods employed by the different authors could contribute to the observed scatter as well. A direct, more homogeneous and systematic spectral analysis of the WAs, similar to what we have done for the UFOs in Paper I and Paper II, would reduce the importance of these possible systematics, but this is beyond the scope of the present paper. Here we focus more on the global picture, noting also that very detailed spectral analysis of WAs could be performed only on a handful of sources as they require very large and high quality datasets. 

Finally, here we tested for simple linear relations in log-space between the absorber parameters, such as ionization, column density, outflow velocity, distance and energetics. This represents a good 1st order approximation but there might be slight variations in the slope, particularly at the low and high ends of the distributions, which could indicate different regimes. This possibility will be briefly considered later in the discussion section.

\section[]{Discussion}

\subsection[]{Unification of the X-ray outflows}

In all the tests considered and in Figures 1--3 we find that the WAs, non-UFOs and UFOs, within the observational uncertainties, show similar relations between their parameters. In particular, the WAs and UFOs lie always at the two ends of the distributions, with the intermediate non-UFOs in between, and they roughly uniformly cover the whole parameter space. When considered all together, we find very significant correlations between their parameters, such as ionization, column density, velocity and distance. These results strongly suggest that these absorbers, sometimes considered of different type, could actually represent parts of a single general stratified outflow observed at different locations along the line of sight. If they were unrelated, the points relative to the different absorbers would be mixed and would not display any significant correlations, especially when considering several sources together. A simple schematic diagram of a stratified accretion disc wind is shown in Fig.~5.  

The stratification along the line of sight can easily explain the relations in Fig.~1--3. For instance, the fact that the faster absorbers are also more ionized suggests that we see components that are ejected closer to the black hole. The increase in column density for higher ionization can be explained with a negative gradient with distance, as also discussed in next section when estimating the density profile. 

The observed correlation between the velocity and the distance shown in Fig.~3c and reported in Table~3 is consistent with an approximate slope of $-0.40\pm0.03$. This is essentially equivalent to the one expected for a wind that has an outflow velocity proportional to the escape velocity at a given location and, therefore, that is able to escape the system and not fall back. Thus, $v_\mathrm{out} = f_\mathrm{v}v_\mathrm{esc}$, where $v_\mathrm{esc} \equiv \left(2GM/r\right)^{1/2}$ is the escape velocity and $f_\mathrm{v} \ge 1$. Therefore, such type of winds are expected to roughly follow a relation $v_\mathrm{out} \propto r^{-1/2}$.

This relation is satisfied in two circumstances: (i) the wind observed at each radius was launched relatively close to that radius with a velocity proportional the local escape speed or (ii) the wind was launched at small radii with a velocity close to the escape speed and then decelerated under the sole effect of gravity, i.e. ballistically. This latter case is probably unphysical because the wind is not expanding in pure vacuum, but there is always some interstellar medium.  

In this regard, the observed relation excludes the scenario of a wind being launched from small radii with a constant terminal velocity, which would lead to a flat profile with radius. However, this last hypothesis is, at least partially, ruled out unless some decelerating processes are present, such as shocks or entrainment of some of the surrounding material. Actually, the fact that the observed profile is $\simeq$$-0.4$ instead of exactly $-0.5$ suggests that some form of interaction with the external medium might be present at $\gg$pc scales. This is supported by the slight increase of the mass outflow rate in Fig.~3d at large distances of $\ga$100pc, which might indicate some entrainment of surrounding material. This point will be addressed also in \S5.4.

As already discussed in Paper III, the location and characteristics of the UFOs are indeed consistent with accretion disc winds/outflows and the possible direct connection with the other non-UFOs and WAs reported here indicates that actually all of them could be identified with the same global outflow observed at different locations along the line of sight, the WAs being ejected at larger distances, of the order of the outer disc and/or torus, the latter possibly representing a natural large-scale extension of the disc itself. (see Fig.~5). In fact, the correlation analysis and the plots in Figures~1--3 points toward an underlying connection among these ionized absorbers. 

It will be discussed later in \S5.3.2 that case (i), in which the wind is launched on a wide range of radii with a velocity profile proportional to the escape velocity, is probably the preferred interpretation and it is directly predicted by stratified MHD accretion disc wind models. This interpretation is valid at least for distances $\ll$100pc scales, as additional effects of interaction with the surrounding medium might be important beyond that (see \S5.4).

Regarding the WAs, we note that the fact that the line of sight projected location of some of them is of the order of the putative obscuring torus at the base of the type 1/type 2 unification models does not necessarily mean that they are produced there, for instance as inferred by Blustin et al.~(2005). The putative torus is also a large, equatorial structure which is difficult to relate with winds that are observed at relatively small inclinations along the line of sight ($\sim$30$^{\circ}$), as for the Seyfert 1 galaxies discussed here (Wu \& Han 2001). We also note that some authors even suggested that the torus itself could actually be identified with the slower and outer regions of a global stratified wind. At these locations, dust would be present at the disc surface and it could be uplifted and become embedded in the outflow, explaining its presence in some WA observations (K\"{o}nigl \& Kartje 1994; Elvis 2000; Kazanas et al.~2012).

Finally, we note that the presence of a photo-ionized outflow extending from the inner regions around the black hole up to $\sim$kpc scales could be directly related with the ionization cones observed through emission lines and images in several Seyfert galaxies, some of them being also part of this sample (e.g. Storchi-Bergmann et al.~2009; Crenshaw \& Kraemer 2000; Dadina et al.~2010; Wang et al.~2011a; Paggi et al.~2012).

\subsection[]{Outflow density profile}

An important quantity describing these ionized outflows is their radial density profile. Considering all these absorbers as different representations of the same phenomenon, we can derive an estimate of their average global density profile of the form $n(r)\propto r^{- \alpha}$.

From the relation between the column density and distance shown in Fig.~3b and reported in Table~3 we have $N_\mathrm{H} = 10^{b} (r/r_\mathrm{s})^{a}$~cm$^{-2}$, where $a = -0.44 \pm 0.04$ and $b = 24.21 \pm 0.20$. The column density can be expressed also as $N_\mathrm{H} = n(r) \Delta r \simeq n(r)r_\mathrm{s}(r/r_\mathrm{s})$, with $\Delta r \sim r$. Combining these two equations we obtain a rough expression for the density profile as $n(r) \simeq n_0 (r/r_\mathrm{s})^{a-1}$, with $n_0 = 10^b/r_\mathrm{s} \simeq (10^{24.21}/3\times 10^5)(M_\mathrm{BH}/M_{\odot})^{-1} \simeq 5\times 10^{10} M_8^{-1}$~cm$^{-3}$ representing the density normalization and $M_8 = M_\mathrm{BH}/10^8M_{\odot}$. Therefore, substituting the observed value of $a\simeq -0.4$, we derive that the density profile has a slope $\alpha = 1 -a \simeq 1.4$. Consequently, for a typical black hole mass of $\sim$$10^7$~$M_{\odot}$ and for an inner radius of log$(r/r_\mathrm{s})$$\simeq$1, we obtain a density at the base of the outflow of $\sim$$10^{10}$~cm$^{-3}$.  

For comparison, from the relation between the column density and the ionization parameter shown in Fig.~1a and following Holczer et al.~(2007) and Behar (2009), an estimate of the radial density profile can be determined using also the absorption measure distribution, defined as $\mathrm{AMD} = d\mathrm{N}_\mathrm{H}/d(\mathrm{log}\xi)$. The AMD is the absorption analog of the emission measure distribution (EMD), widely used in the analysis of the emission line spectra, and it represents the distribution of hydrogen column density along the line of sight as a function of ionization. Given the relation $\mathrm{logN}_\mathrm{H} = a\mathrm{log}\xi + b$ in Table~2 with $a = 0.72 \pm 0.12$, this can then be rewritten as $\mathrm{AMD} = (10^b a \mathrm{ln}10)\xi^a\propto\xi^a$. Then, the slope of the radial density profile can be estimated as $\alpha = (1+2a)/(1+a)$ with uncertainty $\Delta \alpha = \Delta a/(1+a)^2$ (Behar 2009). Substituting the observed quantities in Table~2 we obtain $\alpha = 1.42 \pm 0.04$, consolidating further this important result.

This value is slightly higher than those reported in a more detailed analysis of the WAs in a sample of 5 Seyfert galaxies by Behar (2009), who nevertheless suggested that a slight increase could be present for high ionizations. However, it should also be noted that the possible selection effects for the absorbers with very low and high ionization and column densities previously discussed in \S4.3 could induce a slight steepening of the density profile estimated in this way. Moreover, these estimates do not take into account the possible presence of additional fully ionized material, which does not imprint any observable spectral absorption features. 

Therefore, in both cases the density profile is $n \propto r^{-1.4}$.  As already noted by Behar (2009), this scaling rules out two simple scenarios, a constant density flow ($n \propto r^{0}$) and, on less stringent grounds, a steady, mass conserving spherical symmetric radial flow similar to a stellar wind, in which the mass outflow rate, the opening angle and the wind velocity are all constant ($n \propto r^{-2}$). This density profile suggests that the outflowing gas is more consistent with a conical/paraboloidal shaped wind instead of a simple spherical shell.

\subsection[]{Acceleration mechanisms}

Once we have established the fact that the ionized absorbers can be unified as part of a single, large scale outflow, then a fundamental question follows: what is (are) the main acceleration mechanism(s)? The limited dynamic range in luminosity for the sources considered here hampers a detailed study of possible correlations between the bolometric luminosity and the parameters of the outflows. In fact, the average bolometric luminosity is log$L_\mathrm{bol}$$\simeq$44.5~erg~s$^{-1}$ with a dispersion of only $\simeq$0.8~dex. Moreover, the existence of different absorber components for each source might render the search for such correlations nontrivial. However, we can perform some tests. 

In order to get some insights into the possible acceleration mechanism(s) of the outflow, the velocity alone is not a good parameter to compare with the source luminosity, while the momentum rate and the kinetic power are better because they also take into account the mass outflow rate, i.e. the energetics involved.

Considering the Eddington ratios of the Seyferts in our sample showing X-ray outflows we obtain an average value of $\simeq$0.15 with a dispersion of $\simeq$0.6~dex. This interval is too narrow to investigate for possible correlations/trends between the Eddington ratio and the other parameters. However, we note that for a given ratio, the UFOs are always more powerful than the other outflows.

More insights might be derived comparing the outflow and radiation momentum rates. The outflow momentum rate can be expressed as:

\begin{equation}
\dot{P}_\mathrm{out}= \dot{M}_\mathrm{out}v \simeq 4\pi C_\mathrm{f} m_\mathrm{p} n r^2 v^2,
\end{equation}
where $C_\mathrm{f}$ is the covering fraction and can assume values between 0 and 1. Instead, the momentum rate of the radiation field was defined in \S3 as $\dot{P}_\mathrm{rad}=L_\mathrm{bol}/c$. 
As already introduced in \S4.2 and reported in Fig.~4 and Table~3, there is a significant roughly linear (slope $a = 0.76 \pm 0.19$) correlation between the radiation momentum rate and the outflow momentum rate of the UFOs, therefore $\dot{P}_\mathrm{out} \simeq \dot{P}_\mathrm{rad}$. In particular, the average value of the ratio between these two parameters is consistent with unity, $<\dot{P}_\mathrm{out}/\dot{P}_\mathrm{rad}> = 1.6\pm1.1$, indicating a direct proportionality between these two quantities.

We checked for the existence of a similar relation also for the WAs but we could not constrain it given the large errors and scatter. This is due to the fact that their parameters are less homogeneous and they cover a wider interval of distances compared to the UFOs. Nevertheless, we can estimate a much lower average ratio between their outflow and radiation momentum rates of $\simeq$0.05, with a standard deviation of $\simeq$0.26~dex.

Therefore, the linear correlation observable in Fig.~4 suggests that there should be a significant exchange of momentum between the radiation field and the outflows or, from the relation $\dot{M}_\mathrm{acc} = L_\mathrm{bol}/\eta c^2$, that the power of the outflows is at least related to the accretion rate.

  \begin{figure}
  \centering
   \includegraphics[width=8cm,height=5.5cm,angle=0]{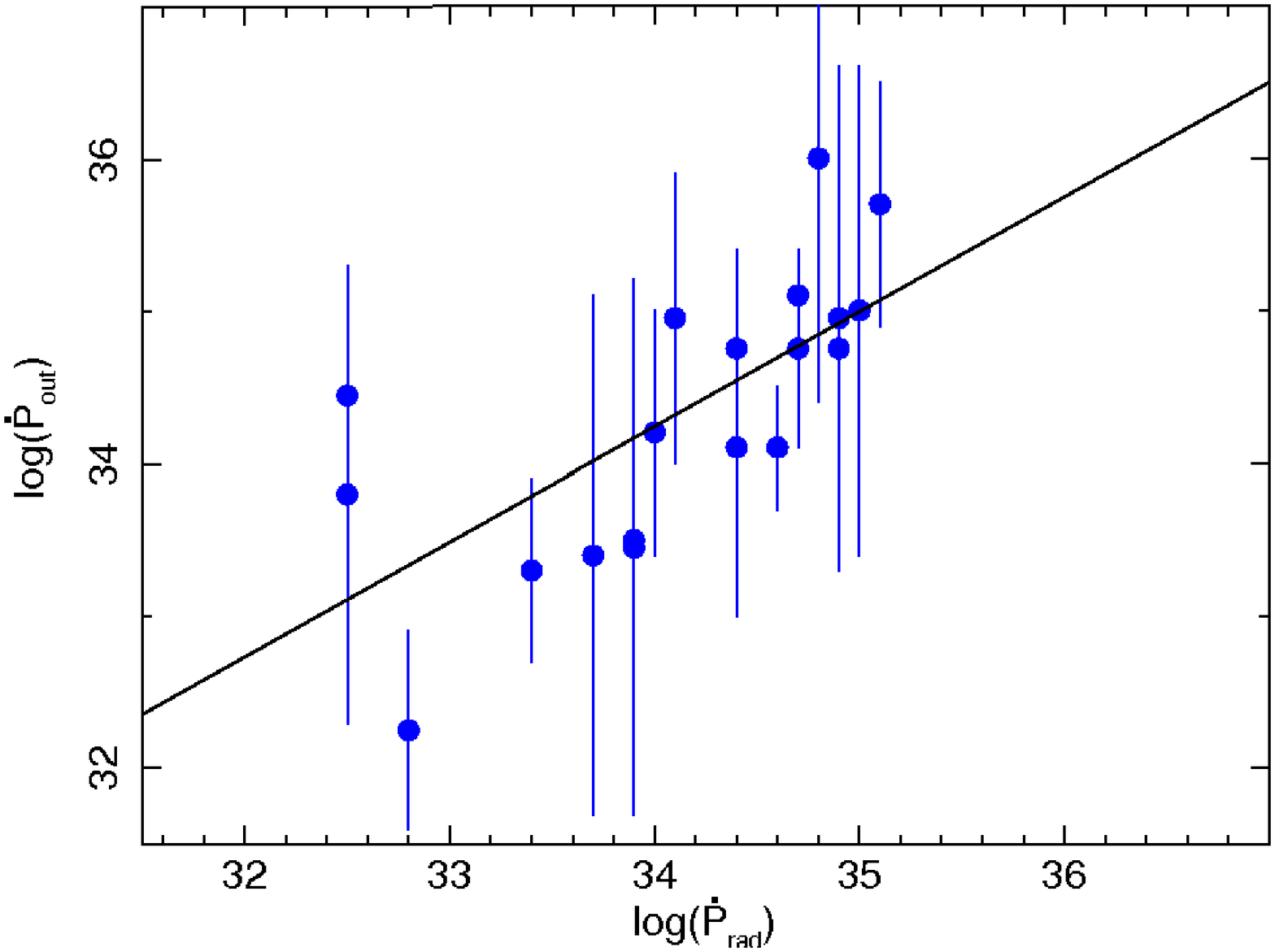}
   \caption{Outflow momentum rate of the UFOs with respect to the momentum rate of the radiation field. The error bars indicate the upper and lower limits and the points are the average between the two. The solid line indicate the best-fit linear regression curve.}
    \end{figure}

\subsubsection[]{Radiation pressure}

As we can see from panel a and b of Fig.~3, even if the inner part of the flow represented by the UFOs can reach significant column densities in the range $N_\mathrm{H} \simeq 10^{22}$--$10^{24}$~cm$^{-2}$, this material is also highly ionized, with possibly the majority of the elements lighter than iron being fully stripped of their electrons. Therefore, it is hard, if not impossible, to invoke radiation pressure from UV absorption lines as the main acceleration mechanism for these flows. This process might instead be more important for winds in bright quasars, given their different SED shape and the higher UV emission compared to Seyferts (e.g., Proga \& Kallman 2004). 

The alternative scenario in which radiation pressure could provide significant acceleration to this material even if it is highly ionized is then through Compton scattering (e.g., King \& Pounds 2003; King 2010a, b). In particular, in this case a direct proportionality between the momentum rate of the radiation field and that of the outflow would be expected:

\begin{equation}
 \dot{P}_\mathrm{out, R} \simeq C_\mathrm{f} \tau_\mathrm{e} \dot{P}_\mathrm{rad},
\end{equation}
where $\tau_\mathrm{e}$ is the electron optical depth to Compton scattering and $C_\mathrm{f}$ is the covering fraction of the wind. If this is the dominating acceleration mechanism, from the linear relation between the UFO momentum flux and that of the radiation field reported in Fig.~4 we derive that the product $C_\mathrm{f} \tau_\mathrm{e}$ should be of the order of unity. 

From the fraction of sources with detected UFOs in Paper I we derive that the global covering fraction of these absorbers is in the range $\sim$0.4--0.6 and therefore we can assume a typical value of $C_\mathrm{f} \simeq 0.5$. Considering the observed average column density of the UFOs of $N_\mathrm{H} \simeq 10^{23}$~cm$^{-2}$ we can estimate an optical depth $\tau_\mathrm{e} \simeq \sigma_\mathrm{T} N_\mathrm{H} \simeq$0.05. The product of these two values is much lower than $\sim$ unity expected from the relation $\dot{P}_\mathrm{out} \simeq \dot{P}_\mathrm{rad}$. However, it should be noted that the column density of the UFOs might have been larger during the acceleration phase. Extrapolating the relation between the column density and the distance shown in Fig.~3b and reported in Table~3 we obtain that the column density at the innermost radii of log$(r/r_\mathrm{s})$$\sim$1 is $N_\mathrm{H} \simeq 10^{24}$~cm$^{-2}$. However, this does not take into account the possibile presence of some additional fully ionized material that is not visible through X-ray spectroscopy. For instance, extrapolating the relation in Fig.~3a down to log$(r/r_\mathrm{s})$$\sim$1 would give rise to log$\xi$$>$5~erg~s$^{-1}$~cm. Therefore, we obtain $\tau_\mathrm{e} \sim 1$, indicating that the material can be mildly Compton-thick at the base of the flow. The product $C_\mathrm{f} \tau_\mathrm{e}$ is now more close to unity. 

Combining equation (5) and (6) and substituting the previously reported definitions for the ionization parameter (\S2), bolometric luminosity and radiation momentum rate (end of \S3), we can derive a rough dependence of the outflow velocity with respect to the ionization parameter:

\begin{equation}
v_\mathrm{out, R} \simeq \left( \frac{k_\mathrm{bol}}{4\pi m_\mathrm{p} c}\right)^{1/2} \tau_\mathrm{e}^{1/2} \xi^{1/2}.
\end{equation} 
therefore, for the radiation pressure case we would expect $v_\mathrm{out} \propto \tau_\mathrm{e}^{1/2} \xi^{1/2}$, which is comparable to what found in Fig.~1b. 

Even if the UFOs have a momentum rate comparable to that of the radiation field, it is important to check if the luminosity of these sources is actually enough to accelerate the material to the escape velocity, which is required for such a wind to leave the system. Therefore, combining equation (7) with the definition of the Eddington luminosity in \S4.2 and imposing that the wind velocity should be equal or higher than the escape velocity at a particular location (see \S5.1), we obtain:

\begin{equation}
\lambda \ga 2 \frac{N_\mathrm{H}}{N_\mathrm{acc}}.
\end{equation}
where $\lambda = L_\mathrm{bol}/L_\mathrm{Edd}$ is the Eddington ratio, $N_\mathrm{H} \equiv n \Delta r \sim n r$ is the observed column density and $N_\mathrm{acc}$ is the column density of the gas during the acceleration phase. From the previously discussed possibility that the column density at the base of the flow (log$(r/r_\mathrm{s})$$\sim$1) , where the majority of the acceleration should take place, is $N_\mathrm{acc} \simeq 10^{24}$~cm$^{-2}$ and considering the average observed column density of the UFOs of $N_\mathrm{H} \simeq 10^{23}$~cm$^{-2}$, we estimate that the radiation is capable to accelerate the wind to the escape velocity if the Eddington ratio is $\ga$0.2. Given that the average Eddington ratio of the sources considered here is only $\lambda$$\simeq$0.15, we suggest that radiation pressure might be relevant to accelerate the observed outflows. 

The effect of radiation pressure could be increased from multiple electron scatterings of the continuum photons if the wind opacity would be higher than one or if the luminosity of these sources would be closer to Eddington. However, the maximum opacity at the very base of the wind is only $\tau_\mathrm{e} \sim 1$ and these sources are sub-Eddington. Regarding this last point, we note that King (2010b) discussed the possibility that bright AGNs, such as those discussed here, could actually be closer to Eddington due to uncertainties on the black hole mass and bolometric luminosity estimates.

Therefore, these evidences indicate that the UFOs may be accelerated, at least partially, by the exchange of momentum with the radiation field through Compton scattering. This is overall consistent with momentum driven outflow models (King \& Pounds 2003; King 2010a, b) which actually predict the existence of highly ionized outflows with velocity $\sim$0.1c and a linear relation between the wind and radiation momentum rates. One requirement by these models is that the optical depth should be of the order of unity, which we find to be possible at the very base of the flow, and that the covering fraction should be $C_\mathrm{f} \sim 1$. From observations we derive that $C_\mathrm{f} \sim 0.5$ (Paper I), which indicates that the outflow has a significant covering fraction and it is uncollimated, nevertheless it is not spherical, but probably more consistent with a conical/paraboloidal bipolar wind-like shape.

\subsubsection[]{Magnetohydrodynamic mechanisms}

An additional mechanism that could provide a concurrent acceleration for this highly ionized material is represented by magnetohydrodynamic (MHD) processes (Blandford \& Payne 1982; Everett \& Ballantyne 2004; Fukumura et al.~2010; Kazanas et al.~2012). We note that MHD mechanisms are a fundamental requirement for generating the viscosity in accretion discs (e.g., Abramowicz \& Fragile 2011) and they are postulated as one of the main heating mechanisms of the X-ray emitting corona (e.g., Haardt \& Maraschi 1991). Moreover, several Seyfert galaxies do show evidence for weak radio jets (Ulvestad \& Wilson 1984; Ulvestad et al.~2005; Wang et al.~2011b). Therefore, significant magnetic fields might well be present in the inner regions of these radio-quiet sources. 

For instance, Fukumura et al.~(2010) studied the ionization structure of self-similar MHD winds off Keplerian accretion discs in AGNs. The magnetic field is dragged by the rotating disc plasma, and as the wind leaves the disc, magnetic torques act on the gas and the wind is magnetocentrifugally accelerated. In this case, the wind is found to end up with a terminal velocity roughly a few times the initial rotational speed. Therefore, a typical feature of these winds is that they retain a clear information about their launching region. At large distances ($\ga$0.1--1pc) the putative AGN torus may actually represent an extension of the outer accretion disc itself.

Then, Kazanas et al.~(2012) generalized this concept and suggested that MHD winds could actually represent the fundamental structure at the base of the unified model of AGNs. In particular, they derived simple scaling laws for these winds and show that they can reproduce several observed properties of different sources. In particular, the simple schematic diagram of a stratified accretion disc wind shown in Fig.~5 is very similar to their MHD wind represented in their Fig.~8.  
Magnetocentrifugally accelerated winds were also previously studied by other authors, for instance Blandford \& Payne (1982) and K\"{o}nigl \& Kartje (1994), the latter suggesting a similar unification scheme as Kazanas et al.~(2012). 

When observed with a certain inclination, the line of sight intercepts distinct components of this stratified MHD accretion disc wind, each with different velocity, column density, ionization and projected distance. This model predicts that the inner part of the flow should be faster and more ionized, being launched closer to the black hole. The fact that the wind is launched from different parts of the disc can provide an explanation of the large range in observed blue-shifted velocities, ionization and column densities of the X-ray absorbers. In this picture, the UFOs and WAs would represent the inner and outer part of this accretion disc wind, respectively. This is consistent with the relations shown in Fig.~1--3 and with the stratified accretion disc wind diagram shown in Fig.~5. 

These self-similar solutions provide the simplest means of deriving a reasonable MHD wind model and allow to derive scale relations among different parameters to directly compare with observations. A realistic case is definitely more complicated than what currently calculated (see Fukumura et al.~(2010) and Kazanas et al.~(2012) for a detailed discussion of the limitations of these calculations). In particular, these models consider the disc only as a boundary condition, a fully self-consistent treatment should take into account the accretion physics as well. However, they are sophisticated enough to provide at least a 1st order characterization of these outflows and allows to investigate the existence of underlying relations/trends. Fukumura et al.~(2010) focused their attention on winds with a radial density profile of the type $n(r)\propto r^{-\alpha}$, with $\alpha$$=$1. This choice was driven by recent observational results on the AMD of some X-ray warm absorbers (Behar 2009). Instead, Blandford \& Payne (1982) adopted a slightly steeper density profile, $\alpha = 1.5$. On the other hand, K\"{o}nigl \& Kartje (1994) also considered solutions with radial density slopes of $\alpha$$=$1 and 1.5, but the solution with $\alpha$$=$1 was preferred because it was representing the ``minimum energy'' configuration. In \S5.2 we estimated that the general radial density profile of the observed outflows has a slope of roughly $\alpha \simeq 1.4$, which is in between these two cases.  

In particular, it is interesting to note the (qualitative) resemblance of the scaling relations and dynamic ranges between the column density, ionization, velocity and distance of the MHD wind model of Fukumura et al.~(2010) and Kazanas et al.~(2012) (shown in the left panel of their Fig.~5 and Fig.~2, respectively) and our observed relations in Fig.~1 (panels a and b) and Fig.~3c. For instance, even if they did not consider the case of the UFOs in their figure, we note the wide range in ionization log$\xi$$\sim$$-1$--5~erg~s$^{-1}$~cm, velocity $v_\mathrm{out}$$\sim$10--10,000~km~s$^{-1}$ and distance log$(r/r_\mathrm{s})$$\sim$4--10. They also find the same trends of increasing ionization, column density and outflow velocity of the absorbers going from large to the small distances to the central black hole.

  \begin{figure}
  \centering
   \includegraphics[width=8cm,height=6cm,angle=0]{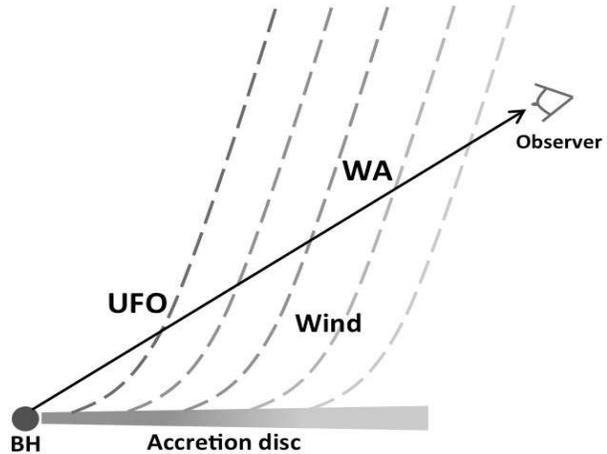}
   \caption{Simple schematic diagram of a stratified accretion disc wind. The figure is not to scale. This is similar to Fig.~8 of Kazanas et al.~(2012). The torus, not directly outlined here, may represent an extension of the outer accretion disc itself.}
    \end{figure}

Therefore, as already derived for the radiative case in \S5.3.2, it is interesting to investigate the possible interpretation as MHD winds using some more quantitative, although still somewhat crude, considerations. The terminal speed of magnetocentrifugal winds is proportional to the Keplerian velocity at the launching region and it is a few times the escape velocity, $v_\mathrm{out}=\omega v_\mathrm{esc}$ where $\omega$ is a factor of the order of $\sim$2--3 and it is the ratio between the Alfv\'en radius and the wind launching radius (Pudritz et al.~2007). The outflow velocity scales with the Keplerian speed as a function of radius, so that the flow will have an onion-like layering of velocities, the fastest inside and the slowest outside. As already noted in \S5.1, this velocity profile is directly consistent with the one observed in Fig.~3c and reported in Table~3. 

It is important to note that, since these winds are accelerated by the action of magnetic torques from magnetic fields that are embedded in the accretion disc, there is an intimate connection between the mass loss rate in the wind and the accretion rate onto the black hole. In particular, one of the most profound scaling relations for MHD winds is represented by the link between the accretion and outflow rates, $\dot{M}_\mathrm{acc} \simeq \omega^2 \dot{M}_\mathrm{out}$ (Pudritz et al.~2007; Reynolds 2012). Combining this with the accretion rate expressed as $\dot{M}_\mathrm{acc}=L_\mathrm{bol}/\eta c^2$, where $\eta$ is the radiative efficiency, and the definition of the momentum flux for the radiation and the wind we can derive the relation:

\begin{equation}
\dot{P}_\mathrm{out, MHD}\simeq\frac{\beta}{\omega^2 \eta} \dot{P}_\mathrm{rad},
\end{equation}
where $\beta = v_\mathrm{out}/c$, meaning that for a given velocity there is a proportionality between the wind and radiation momentum rates also for the MHD case.
From equation (9) and the linear relation $\dot{P}_\mathrm{out} \simeq \dot{P}_\mathrm{rad}$ shown in Fig.~4 and discussed in \S5.3 we find that the radiative efficiency should be a few percent for a typical velocity of the UFOs of $\beta \sim 0.1$ in order to explain this proportionality. Combining equation (9) with the expression for the outflow momentum rate (5) and the definition of the ionization parameter we can derive a relation between the velocity and the ionization parameter for the MHD case as: 

\begin{equation}
v_\mathrm{out, MHD} \simeq \frac{1}{4\pi m_\mathrm{p} c^2} \left( \frac{k_\mathrm{bol}}{\eta \omega^2 C_\mathrm{f}} \right) \xi,
\end{equation}
therefore, in this case we would roughly expect a direct proportionality between the velocity and ionization $v_\mathrm{out} \propto \xi/\eta$. As already done for the radiation pressure case, it is important to check the general conditions under which an MHD wind can actually form, imposing that the outflow velocity from equation (10) is equal or higher than the escape velocity (\S5.1). Using the approximation $\xi \simeq L_\mathrm{ion}/N_\mathrm{H}r$, the definition of the Eddington luminosity and the Schwarzschild radius, we obtain: 

\begin{equation}
\lambda \ga 2 \sqrt{\hat{r}} \tau_\mathrm{e}(\hat{r}) \eta \omega^2 C_\mathrm{f}. 
\end{equation}
where $\lambda = L_\mathrm{bol}/L_\mathrm{Edd}$ is the Eddington ratio, $\tau_\mathrm{e}(\hat{r})$ is the Compton scattering optical depth at the radius $\hat{r} = r/r_\mathrm{s}$. Solving this equation for $\eta$ and considering an average $\lambda \simeq 0.15$, an optical depth at the radius log$\hat{r}$$\sim$1 of $\tau_\mathrm{e}(\hat{r})$$\sim$0.1--1 (see \S5.3.1 and Table~3) and the typical values of $\omega$$\simeq$2--3 and $C_\mathrm{f} \simeq 0.5$, we obtain an estimate of the radiative efficiency of the accretion disc of the Seyfert galaxies considered here in order to be able to efficiently generate MHD winds: $\eta \la \lambda/2 \sqrt{\hat{r}} \tau_\mathrm{e}(\hat{r}) \omega^2 C_\mathrm{f} \la$0.05--0.1. This value of the radiative efficiency is comparable to the typical one derived for quasars of $\eta$$\sim$0.1 (Soltan 1982; Elvis et al.~2002; Davis \& Laor 2011). Regarding Seyfert galaxies in particular, some authors suggested that the radiative efficiency for some of them could be slightly lower than that (Bian \& Zhao 2003b; Panessa et al.~2006; Singh et al.~2011).

Therefore, from these considerations on the wind energetics and the consistency with the expected overall structure/geometry and velocity profile we derive that the observed outflows could be effectively accelerated by MHD processes.

However, it is important to note that this is only a partial conclusion because a complete characterization of these outflows would require the combined treatment of both radiation and MHD effects, both important in AGNs. Some attempts in this direction have been reported in the literature (Proga 2000, 2003; Everett 2005; Ohsuga et al.~2009; Reynolds 2012). If these processes are acting simultaneously, we could naively expect to observe changes between these two regimes. For instance, from equations (7) and (10) we can speculate that the velocity should roughly show a proportionality to $\xi^{1/2}$ from the radiation pressure term and to $\xi$ from the MHD part. It is tempting to compare this with Fig.~1b, where we can see that the residuals of the data with respect to the linear fit are possibly consistent with a similar change of slope between these two regimes going from lower to higher ionization. In fact, the linear regression fit is consistent with an intermediate slope of $\simeq$0.65. Very close to the black hole and for higher ionization/velocities, the MHD regime should always be the dominant part and it might actually enter a regime that is eventually responsible for the acceleration of jets. Further away from the black hole and for lower inization/velocities, radiation forces may contribute more to the properties and dynamics of the flow, depending of the actual state of the material and also the local disc luminosity.

\subsection[]{Energetics and feedback impact}

Having established that the correlations among the different outflow parameters suggest an interpretation as a stratified, large-scale wind with most probably both radiation and MHD mechanisms having a role in the acceleration, then the next step is to, at least qualitatively, investigate the contribution of these outflows on the expected AGN feedback.

An important parameter to quantify the effective feedback contribution of winds/outflows in bright AGNs is the ratio between their mechanical power and the bolometric luminosity. Extensive numerical simulations demonstrated that this value should be around $\sim$5\% and can be as low as $\sim$0.5\% (e.g., Di Matteo et al.~2005; Hopkins \& Elvis 2010). Gaspari et al.~(2011a, b, 2012b) recently demonstrated that outflows with these powers are indeed adequate to produce sufficient feedback to quench cooling flows in elliptical galaxies and to significantly eject gas without overheating the galaxy itself. Since Seyferts are generally hosted in spiral galaxies, which typically have less dense environments compared to ellipticals, then the threshold could be even lower than $\sim$0.1\% (Gaspari et al.~2011a, b, 2012b). 

Simulations of AGN outflows with characteristics equivalent to UFOs have also been independently demonstrated to be able to significantly interact not only with the interstellar medium of the host galaxy but possibly also with the intergalactic medium. They can provide a significant contribution to the quenching of cooling flows and the inflation of bubbles/cavities in the intergalactic medium in both galaxy clusters (e.g., Sternberg et al. 2007; Gaspari et al. 2011a) and especially groups (e.g., Gaspari et al. 2011b). 

In Fig.~6 we plot the kinetic power of the outflows with respect to the bolometric luminosity. We note again that the distribution of points seems rather continuous from the WAs to the UFOs, the latter having a higher power for a given source luminosity. As discussed in Paper III and as evident from Fig.~6, the UFOs have indeed a mechanical power clearly higher than 0.5\% of the bolometric luminosity, with the majority of the values around $\sim$5\%. However, we should note that, as recently reported by Crenshaw \& Kraemer (2012), some of the WAs might actually reach the $\sim$0.5\% level as well when their components are co-added together. Therefore, these outflows, and in particular the UFOs, are clearly the most promising candidates to significantly contribute to the AGN feedback besides radio jets (e.g., Fabian 2012).

  \begin{figure}
  \centering
   \includegraphics[width=5.6cm,height=8.1cm,angle=-90]{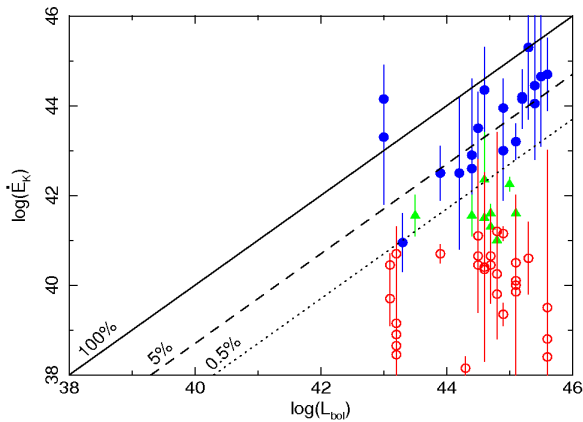}
   \caption{Outflow kinetic power with respect to the bolometric luminosity. The points correspond to the WAs (red open circles), non-UFOs (green filled triangles) and UFOs (blue filled circles), respectively. The error bars indicate the upper and lower limits and the points are the average between the two. The transverse lines indicate the ratios between the outflow mechanical power and bolometric luminosity of 100\% (solid), 5\% (dashed) and 0.5\% (dotted), respectively.}
    \end{figure}

Theoretically, feedback from AGN outflows has been demonstrated to clearly influence the bulge star formation and SMBH growth and possibly also to contribute to the establishment of the observed SMBH-host galaxy relations, such as the $M_\mathrm{BH}$--$\sigma$ (e.g., King 2010a; Ostriker et al.~2010; Power et al.~2011; Zubovas \& King 2012; Faucher-Gigu{\`e}re \& Quataert 2012). Similar and possibly even more massive and/or energetic outflows might have influenced also the formation of structures and galaxy evolution through feedback at higher redshifts, close to the peak of the quasar activity at $z \sim 2$ (Silk \& Rees 1998; Scannapieco \& Oh 2004; Hopkins et al.~2006). 

From Fig.~3a we can note that on $\sim$kpc scales the outflow is very lowly ionized (log$\xi \la 0$~erg~s$^{-1}$~cm) and it could represent the possible conjunction point with large-scale neutral/molecular outflows recently found in some sources in other wavebands (e.g. Nesvadba et al.~2008; Sturm et al.~2011). Several models have been suggested in order to explain their origin, but essentially all of them rely on a two step process in which an initial $\sim$sub-pc scale fast ($v_\mathrm{out} > 1,000$~km~s$^{-1}$) AGN accretion disc outflow perturbs/shocks the interstellar medium, sweeping it up on its way and then decelerates/cools (e.g. King 2010a; Zubovas \& King 2012; Faucher-Gigu{\`e}re \& Quataert 2012). Therefore, it is tempting to check for possible evidences of these effects in our correlation plots. Their intensity should be more prominent at large distances from the black hole. Besides the large uncertainties, in Fig.~3d we can note a slight increase of the mass outflow rate at $>$100pc ($> 10^8 r_\mathrm{s}$) scales, possibly suggesting some entrainment of surrounding material by the wind. In Fig.~3e and f we can also see a slight increase of the wind momentum rate and mechanical power at those locations. These evidences are roughly consistent with the relations reported in Fig.~4 of Faucher-Gigu{\`e}re \& Quataert (2012), who performed a detailed study of the interaction of AGN winds with the surrounding environment and the different regimes of momentum/energy conservation as the resulting shocked material propagates to large distances. 

Observationally, we note that evidences for AGN feedback activity driven by outflows/jets improved significantly in recent years but there are still significant uncertainties, especially regarding the link between the observed phenomenologies at small ($\sim$pc) and large ($\sim$kpc) scales. Promising results on this line have been recently reported for a few Seyfert galaxies, with the detection of bubbles, shocks and jet/cloud interaction, some being also part of our sample (e.g., Wang et al.~2010; Pounds \& Vaughan 2011).

\section{Conclusions}

In order to investigate the possible relations between the ultra-fast outflows (UFOs), mainly detected in the Fe K band through Fe XXV/XXVI absorption lines, and the soft X-ray warm absorbers (WAs) we performed a literature search for papers reporting the analysis of the WAs in the 35 type 1 Seyferts of the sample defined in Paper I. The main results of our study are:

\begin{itemize}

\item The fraction of sources with reported WAs is $>$60\%, consistent with previous studies. The fraction of sources with UFOs is $>$34\%, $>67$\% of which showing also WAs.

\item We reported the main observed WA parameters, such as ionization, column density and outflow velocity. Then, from these values, we estimated also the mass outflow rate, momentum rate and mechanical power. 

\item The large dynamic range obtained when considering all the parameters of these absorbers together allows us, for the first time, to estimate significant correlations among them. We find that the closer the absorber to the black hole, the higher the ionization, column density, velocity and therefore the mechanical power. In particular, in the innermost part of the flow, at distances of log$(r/r_\mathrm{s})$$\simeq$1, we find that the material can be mildly Compton-thick, $N_\mathrm{H} \sim 10^{24}$~cm$^{-2}$, highly ionized, $\mathrm{log}\xi \sim 5$~erg~s$^{-1}$~cm, and the velocity can approach significant fractions of the speed of light.

\item In all the tests the absorber parameters uniformly cover the whole parameter space, with the WAs and UFOs lying always at the two ends of the distribution. This strongly indicates that these absorbers, sometimes considered of different type, could actually be unified in a single, large scale stratified outflow observed at different locations along the line of sight. The UFOs are likely launched from the inner accretion disc and the WAs at larger distances, such as the outer disc and/or torus. See Fig.~5 for a simple schematic diagram of such a stratified wind. 

\item Given the high ionization and velocity of the outflows, and a linear relation between the outflow and radiation momentum rates, we argue that the only two viable acceleration mechanisms are radiation pressure through Compton scattering and MHD processes, the latter playing a major role. In particular, the overall structure/geometry is more consistent with a stratified MHD wind scenario.   

\item Finally, as already discussed in the previous Paper III, here we confirm that these outflows, with the UFOs representing the most energetic part, have a sufficiently high mechanical power ($\ga$0.5\% of the bolometric luminosity) to provide a significant contribution to AGN feedback. In this regard, we find possible evidences for the interaction of the AGN wind with the surrounding environment on large-scales from the correlation plots.

\end{itemize}

In the future, in order to better quantify the predominance of radiation pressure or MHD driving, we should extend the sample to sources with lower and higher Eddington ratios. One would naively expect outflows in lower luminosity AGNs to be dominated by MHD processes while for bright quasars at the other end of the distribution probably radiation pressure is more important. Seyfert galaxies might represent an intermediate case between these two, also suggested by their Eddington ratio of $\sim$0.1. In this respect, we plan also to directly test these hypotheses fitting the data with detailed radiation and MHD wind models (e.g., Sim et al.~2010; Fukumura et al.~2010). 

From an evolutionary point of view, it will be interesting to compare the characteristics of similar outflows found in higher redshift quasars in X-rays (Chartas et al.~2002, 2003, 2009; Giustini et al.~2011; Lanzuisi et al.~2012) and also powerful, large-scale outflows detected in other wavebands (e.g., Nesvadba et al.~2008; Sturm et al.~2011).

Finally, we anticipate that the unprecedented high energy resolution and sensitivity in the wide E$\simeq$0.1--10~keV energy band of the microcalorimeter on board the upcoming \emph{Astro-H} mission will provide important improvements in this field, allowing for the first time to simultaneously study in detail the absorbers in a wide range of ionization states, column densities and velocities.
We also note that the unprecedented effective area of about 10~m$^2$ at 8~keV with even moderate energy resolution of $\simeq$250~eV, such as the one proposed by the \emph{LOFT} mission to the ESA Cosmic Vision, would allow to detect UFOs in bright local AGNs with high significance and at velocities up to $\sim$0.7c thanks to the extended energy bandpass at higher energies.

\section*{Acknowledgments}

The authors thank the anonymous referee for the positive and constructive comments. FT thanks D. Kazanas, K. Fukumura, R.~F. Mushotzky for the useful discussions. MC acknowledges financial support from ASI (contract ASI/INAF/I/009/10/0) and INAF (contract PRIN-INAF-2011). RN was supported by an appointment to the NASA Postdoctoral Program at Goddard Space Flight Center, administered by Oak Ridge Associated Universities through a contract with NASA. This research made use of the StatCodes statistical software hosted by Penn State's Center for Astrostatistics. This research has made use of data obtained from the High Energy Astrophysics Science Archive Research Center (HEASARC), provided by NASA's Goddard Space Flight Center. This research has made use of the NASA/IPAC Extragalactic Database (NED) which is operated by the Jet Propulsion Laboratory, California Institute of Technology, under contract with the National Aeronautics and Space Administration. This research has made use of NASA's Astrophysics Data System.


\begin{thebibliography}{99}
\bibitem[Abramowicz \& Fragile(2011)]{2011arXiv1104.5499A} Abramowicz, M.~A., Fragile, P.~C.\ 2011, arXiv:1104.5499
\bibitem[Akritas \& Bershady(1996)]{1996ApJ...470..706A} Akritas, M.~G., Bershady, M.~A.\ 1996, ApJ, 470, 706
 \bibitem[Akritas \& Siebert(1996)]{1996MNRAS.278..919A} Akritas, M.~G., Siebert, J.\ 1996, MNRAS, 278, 919
\bibitem[Ashton et al.(2006)]{2006MNRAS.366..521A} Ashton, C.~E., Page, M.~J., Branduardi-Raymont, G., Blustin, A.~J.\ 2006, MNRAS, 366, 521 
\bibitem[Behar(2009)]{2009ApJ...703.1346B} Behar, E.\ 2009, ApJ, 703, 1346 
\bibitem[Bian \& Zhao(2003)]{2003MNRAS.343..164B} Bian, W., Zhao, Y.\ 2003a, MNRAS, 343, 164 
\bibitem[Bian \& Zhao(2003)]{2003PASJ...55..599B} Bian, W.~H., Zhao, Y.~H.\ 2003b, PASJ, 55, 599
 \bibitem[Blandford \& Payne(1982)]{1982MNRAS.199..883B} Blandford, R.~D., Payne, D.~G.\ 1982, MNRAS, 199, 883
\bibitem[Blustin et al.(2003)]{2003A&A...403..481B} Blustin, A.~J., Branduardi-Raymont, G., Behar, E., et al.\ 2003, A\&A, 403, 481 
\bibitem[Blustin et al.(2005)]{2005A&A...431..111B} Blustin, A.~J., Page, M.~J., Fuerst, S.~V., Branduardi-Raymont, G., Ashton, C.~E.\ 2005, A\&A, 431, 111
\bibitem[Braito et al.(2007)]{2007ApJ...670..978B} Braito, V., Reeves, J.~N., Dewangan, G.~C., et al.\ 2007, ApJ, 670, 978
\bibitem[Cappi et al.(2009)]{2009A&A...504..401C} Cappi, M., Tombesi, F., Bianchi, S., et al.\ 2009, A\&A, 504, 401
\bibitem[Cardaci et al.(2011)]{2011A&A...530A.125C} Cardaci, M.~V., Santos-Lle{\'o}, M., H{\"a}gele, G.~F., Krongold, Y., D{\'{\i}}az, A.~I., Rodr{\'{\i}}guez-Pascual, P. \ 2011, A\&A, 530, A125 
\bibitem[Chartas et al.(2002)]{2002ApJ...579..169C} Chartas, G., Brandt, W.~N., Gallagher, S.~C., Garmire, G.~P.\ 2002, ApJ, 579, 169
\bibitem[Chartas et al.(2003)]{2003ApJ...595...85C} Chartas, G., Brandt, W.~N., Gallagher, S.~C.\ 2003, Apj, 595, 85
\bibitem[Chartas et al.(2009)]{2009ApJ...706..644C} Chartas, G., Saez, C., Brandt, W.~N., Giustini, M., Garmire, G.~P.\ 2009, ApJ, 706, 644
\bibitem[Costantini et al.(2007)]{2007A&A...461..121C} Costantini, E., Kaastra, J.~S., Arav, N., et al.\ 2007, A\&A, 461, 121 
\bibitem[Crenshaw \& Kraemer(2000)]{2000ApJ...532L.101C} Crenshaw, D.~M., Kraemer, S.~B.\ 2000, ApJ, 532, L101 
\bibitem[Crenshaw \& Kraemer(2012)]{2012ApJ...753...75C} Crenshaw, D.~M., Kraemer, S.~B.\ 2012, ApJ, 753, 75 
\bibitem[Dadina et al.(2005)]{2005A&A...442..461D} Dadina, M., Cappi, M., Malaguti, G., Ponti, G., de Rosa, A.\ 2005, A\&A, 442, 461
\bibitem[Dadina et al.(2010)]{2010A&A...516A...9D} Dadina, M., Guainazzi, M., Cappi, M., Bianchi, S., Vignali, C., Malaguti, G., Comastri, A. \ 2010, A\&A, 516, A9
\bibitem[Dauser et al.(2012)]{2012MNRAS.422.1914D} Dauser, T., Svoboda, J., Schartel, N., et al.\ 2012, MNRAS, 422, 1914
\bibitem[Davis \& Laor(2011)]{2011ApJ...728...98D} Davis, S.~W., Laor, A.\ 2011, ApJ, 728, 98
\bibitem[Di Matteo et al.(2005)]{2005Natur.433..604D} Di Matteo, T., Springel, V., Hernquist, L.\ 2005, Nature, 433, 604 
\bibitem[Elvis(2000)]{2000ApJ...545...63E} Elvis, M.\ 2000, ApJ, 545, 63
\bibitem[Elvis et al.(2002)]{2002ApJ...565L..75E} Elvis, M., Risaliti, G., Zamorani, G.\ 2002, ApJ, 565, L75 
\bibitem[Emmanoulopoulos et al.(2011)]{2011MNRAS.415.1895E} Emmanoulopoulos, D., Papadakis, I.~E., McHardy, I.~M., Nicastro, F., Bianchi, S., Ar{\'e}valo, P. \ 2011, MNRAS, 415, 1895
\bibitem[Everett \& Ballantyne(2004)]{2004ApJ...615L..13E} Everett, J.~E., Ballantyne, D.~R.\ 2004, ApJ, 615, L13 
\bibitem[Everett(2005)]{2005ApJ...631..689E} Everett, J.~E.\ 2005, ApJ, 631, 689
\bibitem[Fabian(2012)]{2012ARA&A..50..455F} Fabian, A.~C.\ 2012, ARA\&A, 50, 455 
\bibitem[Faucher-Gigu{\`e}re \& Quataert(2012)]{2012MNRAS.425..605F} Faucher-Gigu{\`e}re, C.-A., Quataert, E.\ 2012, MNRAS, 425, 605
\bibitem[Ferrarese \& Merritt(2000)]{2000ApJ...539L...9F} Ferrarese, L., Merritt, D.\ 2000, ApJL, 539, L9
\bibitem[Fukumura et al.(2010)]{2010ApJ...715..636F} Fukumura, K., Kazanas, D., Contopoulos, I., Behar, E.\ 2010, ApJ, 715, 636
\bibitem[Gallo et al.(2011)]{2011MNRAS.411..607G} Gallo, L.~C., Miniutti, G., Miller, J.~M., Brenneman, L.~W., Fabian, A.~C., Guainazzi, M., Reynolds, C.~S. \ 2011, MNRAS, 411, 607
\bibitem[Gaspari et al. (2011a)]{} Gaspari, M., Melioli, C., Brighenti, F., D'Ercole, A., 2011a, MNRAS, 411, 349 
\bibitem[Gaspari et al. (2011b)]{} Gaspari, M., Brighenti, F., D'Ercole, A., Melioli, C., 2011b, MNRAS, 415, 1549
\bibitem[Gaspari et al. (2012a)]{} Gaspari, M., Ruszkowski, M., Sharma, P., 2012a, ApJ, 746, 94 
\bibitem[Gaspari et al. (2012a)]{} Gaspari, M., Brighenti, F., Temi P.: 2012b, MNRAS, 424, 190
\bibitem[George et al.(1998)]{1998ApJS..114...73G} George, I.~M., Turner, T.~J., Netzer, H., Nandra, K., Mushotzky, R.~F., Yaqoob, T. \ 1998, ApJS, 114, 73
\bibitem[Ghisellini et al.(2004)]{2004A&A...413..535G} Ghisellini, G., Haardt, F., Matt, G.\ 2004, A\&A, 413, 535
\bibitem[Giustini et al.(2011)]{2011A&A...536A..49G} Giustini, M., Cappi, M., Chartas, G., et al.\ 2011, A\&A, 536, A49
\bibitem[Gofford et al.(2011)]{2011MNRAS.414.3307G} Gofford, J., Reeves, J.~N., Turner, T.~J., et al.\ 2011, MNRAS, 414, 3307
\bibitem[Gofford et al.(2012)]{} Gofford, J., Reeves, J.~N., Tombesi, F., Braito, V., Turner, T.~J., Miller, L., Cappi, M. \ 2012 arXiv:1211.5810
\bibitem[Gondoin et al.(2002)]{2002A&A...388...74G} Gondoin, P., Orr, A., Lumb, D., Santos-Lleo, M.\ 2002, A\&A, 388, 74 
\bibitem[Haardt \& Maraschi(1991)]{1991ApJ...380L..51H} Haardt, F., Maraschi, L.\ 1991, ApJ, 380, L51 
\bibitem[Halpern(1984)]{1984ApJ...281...90H} Halpern, J.~P.\ 1984, ApJ, 281, 90
\bibitem[Holczer et al.(2007)]{2007ApJ...663..799H} Holczer, T., Behar, E., Kaspi, S.\ 2007, ApJ, 663, 799
\bibitem[Hopkins et al.(2006)]{2006ApJS..163....1H} Hopkins, P.~F., Hernquist, L., Cox, T.~J., Di Matteo, T., Robertson, B., Springel, V. \ 2006, ApJS, 163, 1
\bibitem[Hopkins \& Elvis(2010)]{2010MNRAS.401....7H} Hopkins, P.~F., Elvis, M.\ 2010, MNRAS, 401, 7
\bibitem[Jim{\'e}nez-Bail{\'o}n et al.(2008)]{2008MNRAS.391.1359J} Jim{\'e}nez-Bail{\'o}n, E., Krongold, Y., Bianchi, S., Matt, G., Santos-Lle{\'o}, M., Piconcelli, E., Schartel, N. \ 2008, MNRAS, 391, 1359 
\bibitem[Kaastra et al.(2000)]{2000A&A...354L..83K} Kaastra, J.~S., Mewe, R., Liedahl, D.~A., Komossa, S., Brinkman, A.~C.\ 2000, A\&A, 354, L83
\bibitem[Kaspi et al.(2000)]{2000ApJ...535L..17K} Kaspi, S., Brandt, W.~N., Netzer, H., Sambruna, R., Chartas, G., Garmire, G.~P., Nousek, J.~A. \ 2000, ApJ, 535, L17
\bibitem[Kazanas et al.(2012)]{2012AstRv...7c..92K} Kazanas, D., Fukumura, K., Behar, E., Contopoulos, I., Shrader, C.\ 2012, The Astronomical Review, 7, 030000
\bibitem[King \& Pounds(2003)]{2003MNRAS.345..657K} King, A.~R., Pounds, K.~A.\ 2003, MNRAS, 345, 657
\bibitem[King(2010)]{2010MNRAS.402.1516K} King, A.~R.\ 2010a, MNRAS, 402, 1516 
\bibitem[King(2010)]{2010MNRAS.408L..95K} King, A.~R.\ 2010b, MNRAS, 408, L95 
\bibitem[King et al.(2012)]{2012ApJ...746L..20K} King, A.~L., Miller, J.~M., Raymond, J., et al.\ 2012, ApJ, 746, L20 
\bibitem[Konigl \& Kartje(1994)]{1994ApJ...434..446K} Konigl, A., Kartje, J.~F.\ 1994, ApJ, 434, 446
\bibitem[Kraemer et al.(2005)]{2005ApJ...633..693K} Kraemer, S.~B., George, I.~M., Crenshaw, D.~M., et al.\ 2005, ApJ, 633, 693
\bibitem[Krolik \& Kriss(2001)]{2001ApJ...561..684K} Krolik, J.~H., Kriss, G.~A.\ 2001, ApJ, 561, 684
\bibitem[Krongold et al.(2007)]{2007ApJ...659.1022K} Krongold, Y., Nicastro, F., Elvis, M., Brickhouse, N., Binette, L., Mathur, S., Jim{\'e}nez-Bail{\'o}n, E. \ 2007, ApJ, 659, 1022
\bibitem[Laha et al.(2011)]{2011ApJ...734...75L} Laha, S., Dewangan, G.~C., Kembhavi, A.~K.\ 2011, ApJ, 734, 75 
\bibitem[Lanzuisi et al.(2012)]{2012A&A...544A...2L} Lanzuisi, G., Giustini, M., Cappi, M., Dadina, M., Malaguti, G., Vignali, C., Chartas, G. \ 2012, A\&A, 544, A2
\bibitem[Lobban et al.(2011)]{2011MNRAS.414.1965L} Lobban, A.~P., Reeves, J.~N., Miller, L., et al.\ 2011, MNRAS, 414, 1965 
\bibitem[Longinotti et al.(2007)]{2007A&A...470...73L} Longinotti, A.~L., Bianchi, S., Santos-Lleo, M., et al.\ 2007, A\&A, 470, 73 
\bibitem[Longinotti et al.(2010)]{2010A&A...510A..92L} Longinotti, A.~L., Costantini, E., Petrucci, P.~O., et al.\ 2010, A\&A, 510, A92 
\bibitem[Lusso et al.(2010)]{2010A&A...512A..34L} Lusso, E., Comastri, A., Vignali, C., et al.\ 2010, A\&A, 512, A34 
\bibitem[Markowitz et al.(2006)]{2006ApJ...646..783M} Markowitz, A., Reeves, J.~N., Braito, V.\ 2006, ApJ, 646, 783 
\bibitem[Markowitz(2009)]{2009ApJ...698.1740M} Markowitz, A.\ 2009, ApJ, 698, 1740 
\bibitem[Markowitz et al.(2009)]{2009ApJ...691..922M} Markowitz, A., Reeves, J.~N., George, I.~M., et al.\ 2009, ApJ, 691, 922 
\bibitem[Matt et al.(2006)]{2006A&A...445..451M} Matt, G., Bianchi, S., de Rosa, A., Grandi, P., Perola, G.~C.\ 2006, A\&A, 445, 451 
\bibitem[McKernan et al.(2007)]{2007MNRAS.379.1359M} McKernan, B., Yaqoob, T., Reynolds, C.~S.\ 2007, MNRAS, 379, 1359
\bibitem[Nesvadba et al.(2008)]{2008A&A...491..407N} Nesvadba, N.~P.~H., Lehnert, M.~D., De Breuck, C., Gilbert, A.~M., van Breugel, W.\ 2008, A\&A, 491, 407 
\bibitem[Ohsuga et al.(2009)]{2009PASJ...61L...7O} Ohsuga, K., Mineshige, S., Mori, M., Kato, Y.\ 2009, PASJ, 61, L7
\bibitem[Onken et al.(2003)]{2003ApJ...585..121O} Onken, C.~A., Peterson, B.~M., Dietrich, M., Robinson, A., Salamanca, I.~M.\ 2003, ApJ, 585, 121 
\bibitem[Ostriker et al.(2010)]{2010ApJ...722..642O} Ostriker, J.~P., Choi, E., Ciotti, L., Novak, G.~S., Proga, D.\ 2010, ApJ, 722, 642
\bibitem[Paggi et al.(2012)]{2012ApJ...756...39P} Paggi, A., Wang, J., Fabbiano, G., Elvis, M., Karovska, M.\ 2012, ApJ, 756, 39 
\bibitem[Panessa et al.(2006)]{2006A&A...455..173P} Panessa, F., Bassani, L., Cappi, M., et al.\ 2006, A\&A, 455, 173
\bibitem[Peterson et al.(2004)]{2004ApJ...613..682P} Peterson, B.~M., Ferrarese, L., Gilbert, K.~M., et al.\ 2004, ApJ, 613, 682 
\bibitem[Porquet et al.(2004)]{2004A&A...413..913P} Porquet, D., Kaastra, J.~S., Page, K.~L., et al.\ 2004, A\&A, 413, 913 
\bibitem[Pounds et al.(2003)]{2003MNRAS.345..705P} Pounds, K.~A., Reeves, J.~N., King, A.~R., et al.\ 2003, MNRAS, 345, 705
\bibitem[Pounds et al.(2004)]{2004ApJ...616..696P} Pounds, K.~A., Reeves, J.~N., Page, K.~L., O'Brien, P.~T.\ 2004, ApJ, 616, 696 
\bibitem[Pounds \& Vaughan(2011)]{2011MNRAS.413.1251P} Pounds, K.~A., Vaughan, S.\ 2011, MNRAS, 413, 1251 
\bibitem[Power et al.(2011)]{2011MNRAS.413L.110P} Power, C., Zubovas, K., Nayakshin, S., King, A.~R.\ 2011, MNRAS, 413, L110 
\bibitem[Proga(2000)]{2000ApJ...538..684P} Proga, D.\ 2000, ApJ, 538, 684 
\bibitem[Proga(2003)]{2003ApJ...585..406P} Proga, D.\ 2003, ApJ, 585, 406 
\bibitem[Proga \& Kallman(2004)]{2004ApJ...616..688P} Proga, D., Kallman, T.~R.\ 2004, ApJ, 616, 688
\bibitem[Pudritz et al.(2007)]{2007prpl.conf..277P} Pudritz, R.~E., Ouyed, R., Fendt, C., Brandenburg, A.\ 2007, Protostars and Planets V, 277
\bibitem[Reeves et al.(2001)]{2001A&A...365L.134R} Reeves, J.~N., Turner, M.~J.~L., Pounds, K.~A., et al.\ 2001, A\&A, 365, L134 
\bibitem[Reeves et al.(2009)]{2009ApJ...701..493R} Reeves, J.~N., O'Brien, P.~T., Braito, V., et al.\ 2009, Apj, 701, 493
\bibitem[Revnivtsev et al.(2004)]{2004A&A...418..927R} Revnivtsev, M., Sazonov, S., Jahoda, K., Gilfanov, M.\ 2004, A\&A, 418, 927
\bibitem[Reynolds(1997)]{1997MNRAS.286..513R} Reynolds, C.~S.\ 1997, MNRAS, 286, 513 
\bibitem[Reynolds(2012)]{2012ApJ...759L..15R} Reynolds, C.~S.\ 2012, ApJ, 759, L15 
\bibitem[R{\'o}{\.z}a{\'n}ska et al.(2004)]{2004ApJ...600...96R} R{\'o}{\.z}a{\'n}ska, A., Czery, B., Siemiginowska, A., Dumont, A.-M., Kawaguchi, T.\ 2004, ApJ, 600, 96 
\bibitem[Sani et al.(2011)]{2011MNRAS.413.1479S} Sani, E., Marconi, A., Hunt, L.~K., Risaliti, G.\ 2011, MNRAS, 413, 1479
\bibitem[Scannapieco \& Oh(2004)]{2004ApJ...608...62S} Scannapieco, E., Oh, S.~P.\ 2004, ApJ, 608, 62 
\bibitem[Silk \& Rees(1998)]{1998A&A...331L...1S} Silk, J., Rees, M.~J.\ 1998, A\&A, 331, L1 
\bibitem[Singh et al.(2011)]{2011A&A...533A.128S} Singh, V., Shastri, P., Risaliti, G.\ 2011, A\&A, 533, A128 
\bibitem[Sim et al.(2010)]{2010MNRAS.404.1369S} Sim, S.~A., Miller, L., Long, K.~S., Turner, T.~J., Reeves, J.~N.\ 2010, MNRAS, 404, 1369
\bibitem[Smith et al.(2008)]{2008A&A...490..103S} Smith, R.~A.~N., Page, M.~J., Branduardi-Raymont, G.\ 2008, A\&A, 490, 103 
\bibitem[Soltan(1982)]{1982MNRAS.200..115S} Soltan, A.\ 1982, MNRAS, 200, 115 
\bibitem[Starling et al.(2005)]{2005MNRAS.356..727S} Starling, R.~L.~C., Page, M.~J., Branduardi-Raymont, G., et al.\ 2005, MNRAS, 356, 727 
\bibitem[Steenbrugge et al.(2005)]{2005A&A...432..453S} Steenbrugge, K.~C., Kaastra, J.~S., Sako, M., et al.\ 2005, A\&A, 432, 453 
\bibitem[Sternberg et al.(2007)]{2007ApJ...656L...5S} Sternberg, A., Pizzolato, F., Soker, N.\ 2007, ApJ, 656, L5 
\bibitem[Storchi-Bergmann et al.(2009)]{2009MNRAS.394.1148S} Storchi-Bergmann, T., McGregor, P.~J., Riffel, R.~A., et al.\ 2009, MNRAS, 394, 1148 
\bibitem[Sturm et al.(2011)]{2011ApJ...733L..16S} Sturm, E., Gonz{\'a}lez-Alfonso, E., Veilleux, S., et al.\ 2011, ApJ, 733, L16 
\bibitem[Svoboda et al.(2010)]{2010A&A...512A..62S} Svoboda, J., Guainazzi, M., Karas, V.\ 2010, A\&A, 512, A62 
\bibitem[Tarter et al.(1969)]{1969ApJ...156..943T} Tarter, C.~B., Tucker, W.~H., Salpeter, E.~E.\ 1969, ApJ, 156, 943 
\bibitem[Tombesi et al.(2010)]{2010A&A...521A..57T} Tombesi, F., Cappi, M., Reeves, J.~N., et al.\ 2010a, A\&A, 521, A57
\bibitem[Tombesi et al.(2010)]{2010ApJ...719..700T} Tombesi, F., Sambruna, R.~M., Reeves, J.~N., et al.\ 2010b, ApJ, 719, 700
\bibitem[Tombesi et al.(2011)]{2011ApJ...742...44T} Tombesi, F., Cappi, M., Reeves, J.~N., et al.\ 2011a, ApJ, 742, 44
\bibitem[Tombesi et al.(2011)]{2011MNRAS.418L..89T} Tombesi, F., Sambruna, R.~M., Reeves, J.~N., Reynolds, C.~S., Braito, V.\ 2011b, MNRAS, 418, L89
\bibitem[Tombesi et al.(2012)]{2012MNRAS.422L...1T} Tombesi, F., Cappi, M., Reeves, J.~N.,  Braito, V.\ 2012a, MNRAS, 422, L1
\bibitem[Tombesi et al.(2012)]{2012MNRAS.424..754T} Tombesi, F., Sambruna, R.~M., Marscher, A.~P., et al.\ 2012b, MNRAS, 424, 754
\bibitem[Ulvestad \& Wilson(1984)]{1984ApJ...285..439U} Ulvestad, J.~S., Wilson, A.~S.\ 1984, ApJ, 285, 439 
\bibitem[Ulvestad et al.(2005)]{2005ApJ...621..123U} Ulvestad, J.~S., Antonucci, R.~R.~J., Barvainis, R.\ 2005, ApJ, 621, 123 
\bibitem[Vasudevan \& Fabian(2009)]{2009MNRAS.392.1124V} Vasudevan, R.~V., Fabian, A.~C.\ 2009, MNRAS, 392, 1124
\bibitem[Vaughan et al.(2004)]{2004MNRAS.351..193V} Vaughan, S., Fabian, A.~C., Ballantyne, D.~R., et al.\ 2004, MNRAS, 351, 193 
\bibitem[Wandel \& Mushotzky(1986)]{1986ApJ...306L..61W} Wandel, A., Mushotzky, R.~F.\ 1986, ApJ, 306, L61 
\bibitem[Wang \& Zhang(2007)]{2007ApJ...660.1072W} Wang, J.-M., Zhang, E.-P.\ 2007, ApJ, 660, 1072
\bibitem[Wang et al.(2010)]{2010ApJ...719L.208W} Wang, J., Fabbiano, G., Risaliti, G., et al.\ 2010, ApJ, 719, L208
\bibitem[Wang et al.(2011)]{2011ApJ...742...23W} Wang, J., Fabbiano, G., Elvis, M., et al.\ 2011a, ApJ, 742, 23 
\bibitem[Wang et al.(2011)]{2011ApJ...736...62W} Wang, J., Fabbiano, G., Elvis, M., et al.\ 2011b, ApJ, 736, 62 
\bibitem[Winter(2010)]{2010ApJ...725L.126W} Winter, L.~M.\ 2010, ApJ, 725, L126 
\bibitem[Wu \& Han(2001)]{2001ApJ...561L..59W} Wu, X.-B., Han, J.~L.\ 2001, ApJ, 561, L59
\bibitem[Yaqoob et al.(2003)]{2003ApJ...582..105Y} Yaqoob, T., McKernan, B., Kraemer, S.~B., et al.\ 2003, ApJ, 582, 105 
\bibitem[Zhang et al.(2011)]{2011MNRAS.410.2274Z} Zhang, S.~N., Ji, L., Marshall, H.~L., et al.\ 2011, MNRAS, 410, 2274 
\bibitem[Zubovas \& King(2012)]{2012ApJ...745L..34Z} Zubovas, K., King, A.\ 2012, ApJ, 745, L34 
\end{thebibliography}
\end{document}